# Lattice Strategies for the Dirty Multiple Access Channel [†]

Tal Philosof, Ram Zamir[‡], Uri Erez[††] and Ashish Khisti

### Abstract

A generalization of the Gaussian dirty-paper problem to a multiple access setup is considered. There are two additive interference signals, one known to each transmitter but none to the receiver. The rates achievable using Costa's strategies (i.e. by a random binning scheme induced by Costa's auxiliary random variables) vanish in the limit when the interference signals are strong. In contrast, it is shown that lattice strategies ("lattice precoding") can achieve positive rates independent of the interferences, and in fact in some cases - which depend on the noise variance and power constraints - they are optimal. In particular, lattice strategies are optimal in the limit of high SNR. It is also shown that the gap between the achievable rate region and the capacity region is at most 0.167 bit. Thus, the dirty MAC is another instance of a network setup, like the Korner-Marton modulo-two sum problem, where linear coding is potentially better than random binning. Lattice transmission schemes and conditions for optimality for the asymmetric case, where there is only one interference which is known to one of the users (who serves as a "helper" to the other user), and for the "common interference" case are also derived. In the former case the gap between the helper achievable rate and its capacity is at most 0.085 bit.

## I. Introduction

We consider a two-user Gaussian multiple access channel (MAC) with two known interferences as shown in Fig. 1. The interference signals $S_1$ and $S_2$ are known non-causally to the transmitters of user 1 and user 2, respectively, but unknown to the receiver [1]. We consider "*strong interferences*" which are either *arbitrary*, or equivalently, independent Gaussian processes with variances going to infinity. Specifically, we consider the following *doubly dirty MAC* model

$$Y = X_1 + X_2 + S_1 + S_2 + Z, \tag{1}$$

where $Z$ is additive white Gaussian noise, and user 1 and user 2 must satisfy the power constraints $P_1$ and $P_2$, respectively, as shown in Fig. 1.

[†]The material in this paper was presented in part at International Symposium on Information Theory (ISIT) Nice, France, June 2007.

[‡]This work was supported in part by BSF under Grant 2004398.

[††]This work was supported by ISF under Grant 1234/08.

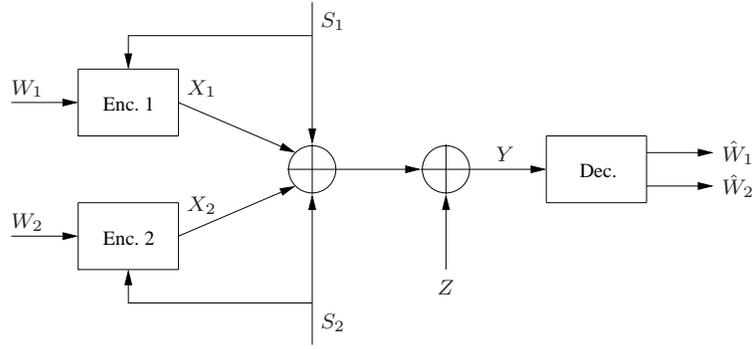

Fig. 1: Doubly dirty MAC.

This channel model generalizes Costa's dirty-paper channel [2] to a multiple access setup. In [2], Costa considered the single-user case, $Y = X + S + Z$, where the interference is assumed to be i.i.d. Gaussian, i.e., $S \sim \mathcal{N}(0, Q)$. It was shown in [2] that in this case, the capacity is $\frac{1}{2}\log_2(1 + \text{SNR})$, where $\text{SNR} = \text{P}/\text{N}$, i.e., as if there was no interference.

The proof of Costa [2] uses the general capacity formula derived by Gel'fand and Pinsker [3] for channels with (non-causal) side information at the transmitter. Their technique falls in the framework of *random binning* which is widely used in the analysis of multi-terminal source and channel coding problems. Using random binning for the direct coding theorem, they obtained a general capacity expression (originally derived for the case of a DMC) which involves an auxiliary random variable $U$:

$$C_{GP} = \max_{p(u,x|s)} \{I(U;Y) - I(U;S)\} \tag{2}$$

where the maximization is over all joint distributions of the form $p(u, s, y, x) = p(s)p(u, x|s)p(y|x, s)$. Selecting the auxiliary random variable $U$ to be

$$U = X + \alpha S, \tag{3}$$

where $X \sim \mathcal{N}(0, P)$ is independent of $S$, and taking $\alpha = \frac{P}{P+N}$, maximizes (2), and the associated random binning scheme is capacity achieving.

Another special case of the channel model (1) was considered by Gel'fand and Pinsker in [4]. They showed that in the noiseless case ($N = 0$), arbitrary large rate pairs $(R_1, R_2)$ are achievable. For the general ($N > 0$) case and independent Gaussian interferences, they conjectured that the capacity region is the same as that of a "clean" MAC, i.e., the standard Gaussian MAC with no interference. The outer bound in Section III below shows that the capacity region is in fact smaller.

An interesting observation we make in this paper is that, in the limit of strong interference, the "natural" generalization of Costa's solution (3) to the doubly dirty MAC is *not* able to achieve *positive rates* (see Section V). In contrast, one dimensional "lattice strategies" [5] can achieve positive rates, and higher rates may be achieved by using high dimensional lattice strategies. Under certain conditions (e.g., high SNR) we show that lattice strategies are in fact asymptotically optimal, i.e., capacity achieving, in the limit of high lattice dimension. Thus, this coding problem is an instance where linear codes are superior to any known random binning technique, see [6] for extensive discussion on this issue. A similar situation was observed by Korner and Marton [7] in a distributed lossless source coding problem, where they showed that the rate region achievable using linear codes is optimal, and is superior to the "best known single letter characterization" for the rate region.

We also consider the case where $S_2 = 0$, i.e., the MAC with a single dirty user (or MAC with only one informed encoder), which knows the interference non-causally at its transmitter. We provide an outer bound and inner bound for this case. The inner bound is based on lattice strategies transmission scheme which are optimal at high SNR. It should be pointed out that for general interferences $S_1$ and $S_2$ in (1) (which are not restricted to be "strong"), the MAC with a single dirty user is a special case of the doubly dirty MAC (by setting $S_2 = 0$). However when we consider the case of strong interferences, the MAC with a single dirty user is not a degeneration of the doubly dirty MAC. The MAC with a single dirty user was considered recently by Somekh-Baruch et al. [8] and Kotagiri and Laneman [9]. The common message ($W_1 = W_2$) capacity of this channel was derived in [8], using generalized random binning by the informed user.

The Gaussian MAC with common interference completes the general framework of MAC problems with full / partial non-causal knowledge of the interference at the encoders. The Gaussian MAC with common interference was considered by Gel'fand and Pinsker [4], and also by Kim et al. [10]. It was shown that like in the point-to-point writing on dirty paper problem, the capacity region of the MAC with common interference is the same as that of the interference-free Gaussian MAC (*clean* MAC). We apply the lattice strategies approach to achieve the capacity region for this case too.

The paper is organized as follows. In Section II, we define the doubly dirty MAC, the MAC with a single dirty user and the MAC with common interference. In Section III we derive outer bounds for the capacity region of the doubly dirty MAC and for the MAC with a single dirty user. A brief review on lattice codes, and a canonical lattice based transmission are presented in Section IV. In Section V we present our results for the doubly dirty MAC. In Section VI we present our results for the MAC with a single dirty user. In Section VII we present a lattice strategies based scheme for the MAC with common interference. Section VIII considers some extensions of these problems. Finally, Section IX concludes the paper.

## II. PROBLEM FORMULATION

### A. The Discrete Memoryless Model

The channel model in (1) is a special case of discrete memoryless MAC with two channel states $S_1 \in \mathcal{S}_1$ and $S_2 \in \mathcal{S}_2$, which are known non-causally at the transmitters of user 1 and user 2, respectively. The states $S_1$ and $S_2$ are memoryless independent states with distributions $p(s_1)$ and $p(s_2)$, respectively. The channel transition probability is $p(y|x_1, x_2, s_1, s_2)$, where $X_1 \in \mathcal{X}_1$ and $X_2 \in \mathcal{X}_2$ are the channel inputs with cardinalities $|\mathcal{X}_1|$ and $|\mathcal{X}_2|$, respectively, and $Y \in \mathcal{Y}$ is the channel output with cardinality $|\mathcal{Y}|$. The channel is memoryless ,i.e,

$$p(\mathbf{y}|\mathbf{x}_1, \mathbf{x}_2, \mathbf{s}_1, \mathbf{s}_2) = \prod_{i=1}^{n} p(y_i|x_{1i}, x_{2i}, s_{1i}, s_{2i}), \tag{4}$$

where bold face indicates $n$-tuples vectors. The encoder outputs of user 1 and user 2 are given by

$$\mathbf{x}_i = f_i(w_i, \mathbf{s}_i) \quad \text{for}, i = 1, 2,$$

where $w_i \in \mathcal{W}_i$ are the transmitted messages. The achievable rates are indicated by $R_1$ and $R_2$ where $|\mathcal{W}_1| = 2^{nR_1}$ and $|\mathcal{W}_2| = 2^{nR_2}$. The decoder reconstructs the transmitted messages $w_1, w_2$ from the channel output, hence

$$(\hat{w}_1, \hat{w}_2) = g(\mathbf{y}).$$

A single letter characterization for the capacity region is not known; see [6], [11] for a more detailed discussion. The best known achievable rate region for this channel, based on the random binning technique of [12], is given by the convex hull of all rate pairs $(R_1, R_2)$ satisfying

$$R_1 \leq I(U_1; Y|U_2) - I(U_1; S_1)$$
$$R_2 \leq I(U_2; Y|U_1) - I(U_2; S_2) \tag{5}$$
$$R_1 + R_2 \leq I(U_1, U_2; Y) - I(U_1; S_1) - I(U_2; S_2).$$

for some $p(u_1, u_2, x_1, x_2|s_1, s_2) = p(u_1, x_1|s_1)p(u_2, x_2|s_2)$, where $|\mathcal{U}_i| \leq |\mathcal{X}_i| + |\mathcal{S}_i|$ for $i = 1, 2$.

### B. The Gaussian Model

In this paper we focus on the Gaussian channel case. Specifically we consider the following models:

*1) Doubly dirty MAC:*

$$Y = X_1 + X_2 + S_1 + S_2 + Z, \tag{6}$$

where $Z \sim \mathcal{N}(0, N)$ is independent of $X_1, X_2, S_1, S_2$, and where user 1 and user 2 must satisfy the power constraints $\frac{1}{n}\sum_{i=1}^{n} x_{1,i}^2 \leq P_1$ and $\frac{1}{n}\sum_{i=1}^{n} x_{2,i}^2 \leq P_2$ respectively, see Fig. 1. The interferences $S_1$ and $S_2$ are known non-causally to the transmitters of user 1 and user 2, respectively. We define the signal-to-noise ratio for each user as $\text{SNR}_1 = \frac{P_1}{N}$ and $\text{SNR}_2 = \frac{P_2}{N}$. In this paper we focus on the *strong interferences* case which are either *arbitrary*, or equivalently independent Gaussian with variances going to infinity.

*2) MAC with a single dirty user and the "helper problem":*

$$Y = X_1 + X_2 + S_1 + Z \tag{7}$$

In this asymmetric case, user 1 knows the interference $S_1$ (informed user) and user 2 is not aware of the interference (uninformed user), as shown in Fig. 2. The "helper problem" is a special case where the uninformed user tries to send information at his maximal rate, while the sole role of the informed user is to help the uninformed user and it does not send any information.

As indicated in the Introduction, the dirty MAC with a single interference may at first glance appear as a special case of doubly dirty MAC for $S_2 = 0$; it is the case if we consider general interferences $S_1$ and $S_2$ (which are not restricted to be strong interferences) in (6). In this paper, we consider the strong interferences case, hence the MAC with a single (strong) dirty user is not a special case of the doubly dirty MAC with strong interferences.

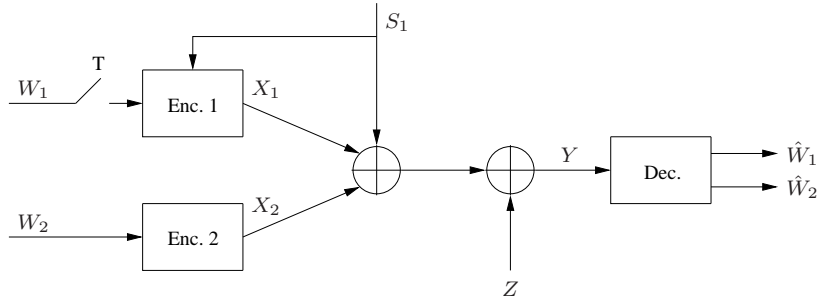

Fig. 2: MAC with a single dirty user (when $T$ is off: the helper problem).

*3) MAC with common Interference:* In this case there is a single interference $S_c$ which is known non-causally to both encoders as shown in Fig. 3, i.e.,

$$Y = X_1 + X_2 + S_c + Z, \tag{8}$$

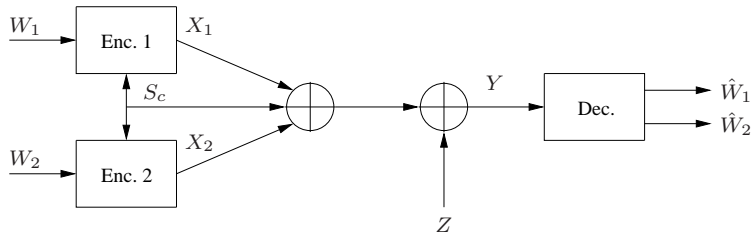

Fig. 3: MAC with common interference.

## III. Outer Bounds for the Gaussian Dirty MAC

We first establish an outer bound on the capacity region of the MAC with a single dirty user (7), and then we use this result to get an outer bound for the doubly dirty MAC (6) for *strong interferences*; where the term "strong interferences" refers to *arbitrary* interference sequences or the limit of independent Gaussian interferences with variances going to infinity.

**Theorem 1 (single dirty user outer bound).** *In the limit of strong interference, the capacity region of the MAC*

*with a single dirty user (user 1)* (7) *is contained in the following region:*

$$R_2 \leq \frac{1}{2} \log_2 \left( 1 + \frac{\min\{P_1, P_2\}}{N} \right)$$

$$R_1 + R_2 \leq \frac{1}{2} \log_2 \left( 1 + \frac{P_1}{N} \right)$$

(9)

*Proof:* For simplicity, we first derive the outer bound for Gaussian $S_1$ with variance $Q_1$, i.e., $S_1 \sim \mathcal{N}(0, Q_1)$ where $Q_1 \to \infty$.

Assume that a genie reveals the message of user 1 to user 2 and vice versa, implying that, in fact, both users intend to transmit a common message $W$. An upper bound on the rate of this message clearly upper bounds $R_1 + R_2$ for the independent messages case ($W_1 \neq W_2$). Applying Fano's inequality to the common message rate $R$ we have,

$$nR \leq H(W) = H(W|Y^n) + I(W; Y^n) \leq n\epsilon_n + I(W; Y^n),$$

where $\epsilon_n \to 0$ as the error probability ($P_e^{(n)}$) goes to zero. The following chain of inequalities can be easily verified.

$$I(W; Y^n) = h(Y^n) - h(Y^n|W)$$

$$\leq h(Y^n) - h(Y^n|W, X_2^n) \tag{10}$$

$$= h(Y^n) - h(Y^n|W, X_2^n, S_1^n) - I(S_1^n; Y^n|W, X_2^n) \tag{11}$$

$$\leq h(Y^n) - h(Z^n) - I(S_1^n; Y^n|W, X_2^n) \tag{12}$$

$$= h(Y^n) - h(Z^n) - h(S_1^n) + h(S_1^n|W, X_2^n, Y^n) \tag{13}$$

$$= h(Y^n) - h(Z^n) - h(S_1^n) + h(X_1^n + Z^n|W, X_2^n, Y^n) \tag{14}$$

$$\leq h(Y^n) - h(Z^n) - h(S_1^n) + h(X_1^n + Z^n), \tag{15}$$

where the equality in (13) follows from the fact that $S_1^n$ is independent of $(X_2^n, W)$ and the two inequalities are a consequence of the fact that conditioning reduces differential entropy. Since $S_1 \sim \mathcal{N}(0, Q_1)$ where $Q_1 \to \infty$, we have by the Cauchy-Schwartz inequality $h(Y^n) \leq \frac{n}{2} \log_2 2\pi e(N + (\sqrt{P_1} + \sqrt{P_2} + \sqrt{Q_1})^2) = n \left( \frac{1}{2} \log_2 Q_1 + o(1) \right)$ where $o(1) \to 0$ for fixed $P_1$, $P_2$ as $Q_1 \to \infty$, and $h(S_1^n) = \frac{n}{2} \log_2 2\pi e Q_1$. Substituting in (15), and setting $\epsilon_n' = \epsilon_n + o(1)$ we have

$$nR \leq n\epsilon' + h(X_1^n + Z^n) - h(Z^n) \leq \frac{n}{2} \log_2 \left( 1 + \frac{P_1}{N} \right) + n\epsilon_n',$$

as stated in the sum-rate bound in (9). The bound on $R_2$ trivially follows by revealing $S_1^n$ to the decoder.

To complete the proof, we still need to justify that the above outer bound is also agree with any *arbitrary* interference sequence. The set of randomly drawn Gaussian interference sequences with variance going to infinity is contained in the set of arbitrary interference sequences. This implies that the outer bound for the capacity region of the arbitrary interference should also bound from above the capacity region for Gaussian interference with variance going to infinity. $\qquad\square$

The outer bound is depicted in Fig. 4 and in Fig. 5 for $P_1 \leq P_2$ and for $P_1 > P_2$, respectively, where $C(x) \triangleq \frac{1}{2} \cdot \log_2(1+x)$. From Fig. 5, the corner point $(R_1^c, R_2^c)$ is given by

$$R_1^c = \frac{1}{2} \log_2 \left( \frac{P_1 + N}{P_2 + N} \right)$$
$$R_2^c = \frac{1}{2} \log_2 \left( 1 + \frac{P_2}{N} \right). \tag{16}$$

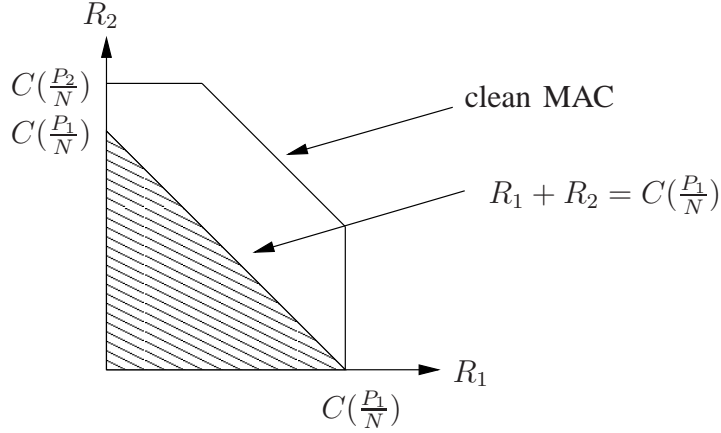

Fig. 4: Outer bound for MAC with a single dirty user (user 1) for $P_1 \leq P_2$.

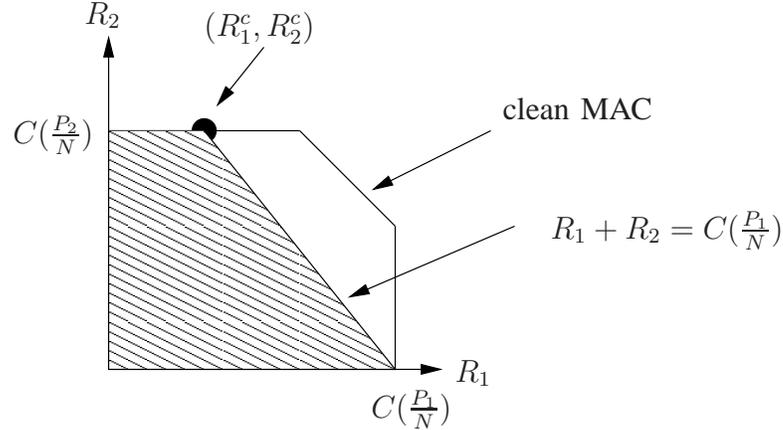

Fig. 5: Outer bound for MAC with a single dirty user (user 1) for $P_1 > P_2$.

The outer bound in Theorem 1 is specialized to the *helper* problem in the following corollary.

**Corollary 1 (helper problem outer bound).** *If only user 2 (the uninformed user) intends to send the message (i.e., $R_1 = 0$) in the single dirty user model (7), then in the limit of strong interference, an upper bound on the rate $R_2$ is given by*

$$R_2 \leq \frac{1}{2} \log_2 \left( 1 + \frac{\min\{P_1, P_2\}}{N} \right). \tag{17}$$

The outer bound (9) for the single dirty user case is also an outer bound for the doubly dirty MAC, provided that the $S_1$ and $S_2$ are independent Gaussian interferences with infinite variances or arbitrary interference sequences. One can show a tighter outer bound by taking the intersection of the outer bounds for MAC with a single interference $S_1$ known to user 1 and MAC with a single interference $S_2$ known to user 2.

**Corollary 2 (doubly dirty MAC outer bound).** *In the limit of strong interference, the capacity region of the doubly dirty MAC* (6) *with $S_1$ and $S_2$ independent is contained in the following region:*

$$R_1 + R_2 \quad \leq \quad \frac{1}{2} \log_2 \left( 1 + \frac{\min\{P_1, P_2\}}{N} \right).$$ (18)

The bound is plotted in Figure 6.

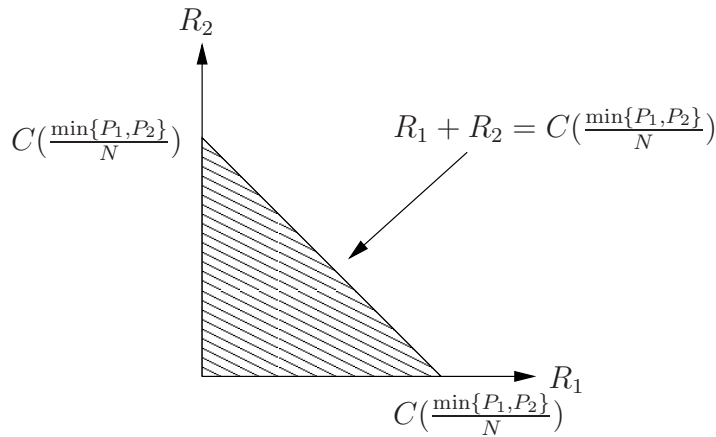

Fig. 6: Outer bound for the doubly dirty MAC in Fig. 1.

The proof of Theorem 1 implies that the upper bound for the rate sum also upper bounds the common message capacity of the doubly dirty MAC, i.e., when both encoders intend to send the same message $W$. In [8] Smoech-Baruch et al. find the common message capacity for the single dirty user case. The outer bound (18) can also be derived from [8] for the case of strong interference.

Gel'fand and Pinsker in [4] showed that in the noiseless case ($N = 0$), arbitrary large rate pairs $(R_1, R_2)$ are achievable. For the general case ($N > 0$) and independent Gaussian interferences, they conjectured that the capacity region is the same as that of the MAC with no interference (clean MAC). The outer bound for the doubly dirty MAC (18) contradicts their conjecture.

## IV. Lattice Based Transmission

### A. Preliminary: Lattices

An $n$-dimensional lattice $\Lambda$ is a discrete group in the Euclidian space $\mathbb{R}^n$ which is closed with the respect to the addition operation (over $\mathbb{R}$) [13]. The lattice is specified by

$$\Lambda = \{\lambda = G \cdot \mathbf{i} \; : \; \mathbf{i} \in \mathbb{Z}^n\},$$ (19)

where $G$ is an $n \times n$ real valued matrix called the lattice generator matrix. A coset of the lattice is any translation of the original lattice $\Lambda$, i.e, $\mathbf{a} + \Lambda$ where $\mathbf{a} \in \mathbb{R}^n$.

The nearest neighbor quantizer $Q_\Lambda(\cdot)$ associated with $\Lambda$ is defined by

$$Q_\Lambda(\mathbf{x}) = \lambda \in \Lambda \quad \text{if } ||\mathbf{x} - \lambda|| \leq ||\mathbf{x} - \lambda'||, \quad \forall \lambda' \in \Lambda, \tag{20}$$

where $|| \cdot ||$ denotes Euclidian norm. The Voronoi region of a lattice point $\lambda$ is the set of all points in $\mathbb{R}^n$ that are closer (in Euclidian distance) to $\lambda$ than to any other lattice point. Specifically, the fundamental Voronoi region is defined as the set of all points that are closest to the origin

$$\mathcal{V} = \{\mathbf{x} \in \mathbb{R}^n : Q_\Lambda(\mathbf{x}) = \mathbf{0}\}, \tag{21}$$

where ties are broken arbitrarily. The modulo lattice operation with respect to $\Lambda$ is defined as

$$\mathbf{x} \bmod \Lambda = \mathbf{x} - Q_\Lambda(\mathbf{x}). \tag{22}$$

The modulo lattice operation satisfies the following distributive property

$$[\mathbf{x} \bmod \Lambda + \mathbf{y}] \bmod \Lambda = [\mathbf{x} + \mathbf{y}] \bmod \Lambda. \tag{23}$$

The second moment of a lattice $\Lambda$ is given by

$$\sigma_\Lambda^2 = \frac{\frac{1}{n} \int_{\mathcal{V}_0} ||\mathbf{x}||^2 \mathbf{dx}}{V}, \tag{24}$$

where $V$ is the volume of the fundamental Voronoi region, i.e., $V = \int_{\mathcal{V}_0} \mathbf{dx}$ (the same for all Voronoi regions of $\Lambda$). The normalized second moment is given by

$$G(\Lambda) = \frac{\sigma_\Lambda^2}{V^{2/n}}. \tag{25}$$

The normalized second moment is always greater than $1/2\pi e$. It is known [14] that for sufficiently large dimension there exist lattices which are good for quantization (these lattices are also known as good lattices for shaping [15]), in the sense that for any $\epsilon > 0$

$$\log_2(2\pi e G(\Lambda)) < \epsilon, \tag{26}$$

for large enough $n$. On the other hand, there exist of lattices with second moment $P$ which are good for AWGN channel coding satisfying [15]

$$\Pr(\mathbf{X} \notin \mathcal{V}) < \epsilon, \text{ where } \mathbf{X} \sim \mathcal{N}(\mathbf{0}, (P - \epsilon)I_n), \ \forall \epsilon > 0, \tag{27}$$

where $I_n$ is an $n \times n$ identity matrix.

The differential entropy of an $n$-dimensional random vector $\mathbf{D}$ which is distributed uniformly over the fundamental Voronoi cell, i.e., $\mathbf{D} \sim \text{Unif}(\mathcal{V})$ is given by [14]

$$h(\mathbf{D}) = \log_2(V) \tag{28}$$

$$= \log_2\left(\frac{\sigma_\Lambda^2}{G(\Lambda)}\right)^{n/2} \tag{29}$$

$$= \frac{n}{2} \log_2\left(\frac{\sigma_\Lambda^2}{G(\Lambda)}\right) \tag{30}$$

$$\approx \frac{n}{2} \log_2\left(2\pi e \sigma_\Lambda^2\right), \tag{31}$$

where the last (approximate) equality holds for lattices which are good for quantization.

*B. A Canonical Lattice Transmission Scheme*

We provide a general lattice-based transmission scheme which will be used for the Gaussian doubly dirty MAC (6) and for the single informed user case (7). The transmission schemes that we use in Section V and Section VI are special cases of the following general transmission scheme, depicted in Fig. 7.

In the following transmission scheme, encoder 1 and encoder 2 use the lattices $\Lambda_1$ and $\Lambda_2$ with second moment $P_1$ and $P_2$ and fundamental Voronoi region $\mathcal{V}_1$ and $\mathcal{V}_2$, respectively. The encoders send the following signals:

$$\mathbf{X}_1 = [\mathbf{V}_1 - \alpha_1 \mathbf{S}_1 + \mathbf{D}_1] \bmod \Lambda_1$$
$$\mathbf{X}_2 = [\mathbf{V}_2 - \alpha_2 \mathbf{S}_2 + \mathbf{D}_2] \bmod \Lambda_2, \tag{32}$$

where $\alpha_1, \alpha_2 \in [0,1]$; $\mathbf{V}_1 \in \mathrm{Unif}(\mathcal{V}_1)$ and $\mathbf{V}_2 \in \mathrm{Unif}(\mathcal{V}_2)$ are independent and carry the information of user 1 and user 2, respectively. The encoders use independent (pseudo-random) dither signals $\mathbf{D}_1 \sim \mathrm{Unif}(\mathcal{V}_1)$ and $\mathbf{D}_2 \sim \mathrm{Unif}(\mathcal{V}_2)$, where $\mathbf{D}_1$ is known to encoder 1 and to the decoder, and $\mathbf{D}_2$ is known to encoder 2 and to the decoder, as shown in Fig. 7. From the dither property [14], $\mathbf{X}_1 \sim \mathrm{Unif}(\mathcal{V}_1)$ independent of $\mathbf{V}_1$ and $\mathbf{X}_2 \sim \mathrm{Unif}(\mathcal{V}_2)$ independent of $\mathbf{V}_2$, hence the power constraints are satisfied.

The decoder reduces modulo-$\Lambda_r$ the term $\alpha_r \mathbf{Y} - \gamma \mathbf{D}_1 - \beta \mathbf{D}_2$, i.e.,

$$\mathbf{Y}' = [\alpha_r \mathbf{Y} - \gamma \mathbf{D}_1 - \beta \mathbf{D}_2] \bmod \Lambda_r. \tag{33}$$

where $\alpha_r \in [0,1]$. The scalars $\alpha_1, \alpha_2, \alpha_r, \beta, \gamma$ and the lattices $\Lambda_1, \Lambda_2, \Lambda_r$ will be chosen differently in each scenario in the sequel.

The main advantage of the lattice based transmission above is its robustness. Unlike in the random binning technique, the achievable rates of the lattice based scheme are ignorant to the exact distributions of the interferences. Hence, this scheme is appropriate for arbitrary interference sequences.

## V. The Gaussian Doubly Dirty MAC

In this section we present lattice strategies based transmission schemes for the Gaussian doubly dirty MAC (6) for high SNR as well as for general SNR as well. We also present conditions for optimality and a uniform bound for the gap from capacity for these schemes.

In the "writing on dirty paper" problem, the capacity can be achieved both by using the random binning technique [2] or using lattice strategies [5]. However, in the Gaussian doubly dirty MAC the random binning technique has inferior achievable rate region with respect to lattice strategies as we will discuss below.

A straightforward generalization of Costa's random binning scheme using the auxiliary random variable given in (3) for the doubly dirty MAC (6) fails to achieve *any positive rates*. To see that, consider for simplicity the limit of high SNR where $\mathrm{SNR}_1, \mathrm{SNR}_2 \gg 1$, and strong independent Gaussian interferences, i.e., $S_1 \sim \mathcal{N}(0, Q_1)$ and $S_2 \sim \mathcal{N}(0, Q_2)$ where $Q_1, Q_2 \gg \max\{P_1, P_2\}$. Clearly from Gel'fand and Pinsker capacity (2) for the point-to-point case, for any $U_1, U_2$ (with any joint distribution) the rate sum of the doubly dirty MAC is upper bounded by $h(U_1, U_2|S_1, S_2) - h(U_1, U_2|Y)$. Taking the "natural" generalization of (3) to the two users case, the auxiliary random variables are $U_1 = X_1 + S_1$ and $U_2 = X_2 + S_2$ (since $\alpha = \frac{P_1}{P_1+N} \approx 1$ at high SNR), and $X_1 \sim \mathcal{N}(0, P_1)$, $X_2 \sim \mathcal{N}(0, P_2)$ are independent, hence from (5) we get that

$$
\begin{aligned}
R_1 + R_2 &= \big[h(U_1, U_2 | S_1, S_2) - h(U_1, U_2 | Y)\big]^+ \\
&= \big[h(X_1, X_2) - h(Y | U_1, U_2) - h(U_1, U_2) + h(Y)\big]^+ \\
&= \big[h(X_1, X_2) - h(Z) - h(U_1) - h(U_2) + h(Y)\big]^+ \\
&\leq \big[h(S_1 + S_2) - h(S_1) - h(S_2) - h(Z)\big]^+ + o(1) \to 0,
\end{aligned}
$$

as $Q_1, Q_2 \to \infty$, where $[x]^+ \triangleq \max\{x, 0\}$. Therefore, we can not achieve any positive rates using the random binning scheme induced by these random binning variables.

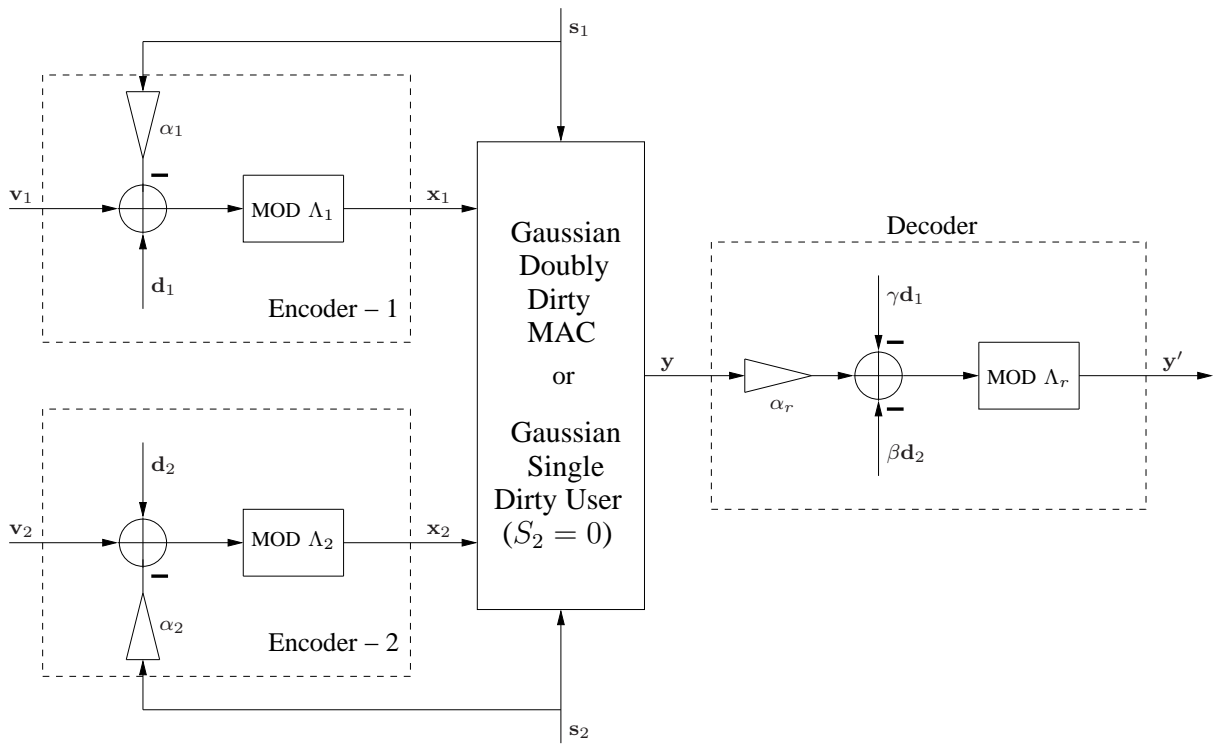

Fig. 7: Canonical transmission scheme.

Nevertheless, inspired by the lattice strategies approach of [5] for the causal dirty paper channel, if we select the auxiliary random variables $U_i$ in (2) to be a periodic function of $X_i$ and $S_i$ for $i = 1, 2$, we get positive rates. Specifically, we choose

$$
U_i = [X_i + \alpha_i S_i + D_i] \bmod \Delta_i, \quad i = 1, 2 \tag{34}
$$

where $\Delta_i = \sqrt{12P_i}$, the transmitted signal is $X_i \sim \text{Unif}\left([-\frac{\Delta_i}{2}, \frac{\Delta_i}{2})\right)$ which is independent of $S_i$ for $i = 1, 2$, and $D_i$ is the dither signal which is uniformly distributed over the interval $[-\frac{\Delta_i}{2}, \frac{\Delta_i}{2})$. The power constraints are satisfied since $P_i = \Delta_i^2 / 12$ for $i = 1, 2$. This choice is equivalent to a one dimensional "lattice strategy" in the same sense of Shannon causal strategies [16], [5]. In Section V-B, we show that the achievable sum-rate using the auxiliary random variables in (34) for the case that $P_1 = P_2 = P$ is bounded from below by

$$R_1 + R_2 \geq u.c.e \left\{ \left[ \frac{1}{2} \log_2 \left( \frac{1}{2} + \frac{P}{N} \right) - \frac{1}{2} \log_2 \left( \frac{\pi e}{6} \right) \right]^+ \right\},$$

where $u.c.e$ is the upper convex envelope with respect to $P/N$.

Similarly to the point-to-point case [5], a natural extension of (34) is to use high dimensional lattice strategies. In the sequel, we show that this technique is in fact optimal for the doubly dirty MAC problem (6) in some cases.

### A. Lattice Strategies at High SNR

We present lattice strategies scheme for the doubly Dirty MAC (6), and show that they are optimal at high SNR, i.e., $\mathrm{SNR}_1, \mathrm{SNR}_2 \gg 1$. The transmission schemes in this section and in Section V-B, which are based on lattice strategies, are known to be optimal for the (single user) "writing on dirty paper" problem [5].

In the following scheme we use an $n$-dimensional lattice $\Lambda$, with normalized second moment $G(\Lambda)$ and fundamental Voronoi region $\mathcal{V}$. We assume that $\Lambda$ has a second moment equal to $\min\{P_1, P_2\}$.

**Theorem 2.** *At high SNR and in the limit of strong interferences, the capacity region of the doubly dirty MAC (6), is given by the set of all rate pairs $(R_1, R_2)$ satisfying*

$$R_1 + R_2 \leq \frac{1}{2} \log_2 \left( \frac{\min\{P_1, P_2\}}{N} \right) - o(1),$$

*where $o(1) \to 0$ as $\min\{P_1, P_2\} \to \infty$.*

*Proof:* The converse follows from Corollary 2. To prove the direct part we use a lattice $\Lambda$ with second moment $\min\{P_1, P_2\}$ which is good for quantization (26). The transmission scheme follows the canonical transmission scheme of Section IV-B with $\Lambda_1 = \Lambda_2 = \Lambda_r = \Lambda$ and $\alpha_1 = \alpha_2 = \alpha_r = 1$, and without dither signals, hence the encoders are given by

$$\mathbf{X}_1 = [\mathbf{V}_1 - \mathbf{S}_1] \bmod \Lambda$$

$$\mathbf{X}_2 = [\mathbf{V}_2 - \mathbf{S}_2] \bmod \Lambda,$$

where $\mathbf{V}_1, \mathbf{V}_2 \sim \mathrm{Unif}(\mathcal{V})$ carry the information for user 1 and user 2, respectively. The transmitted signals $\mathbf{X}_1, \mathbf{X}_2 \sim \mathrm{Unif}(\mathcal{V})$, hence the power constraints are satisfied in the average over all the codewords[1]. The received signal $\mathbf{Y}$ is reduced modulo-$\Lambda$. Using the modulo distributive property (23), we get the following equivalent additive modulo MAC

$$\mathbf{Y}' = \mathbf{Y} \bmod \Lambda = [\mathbf{V}_1 + \mathbf{V}_2 + \mathbf{Z}] \bmod \Lambda.$$

The achievable rate sum over this modulo MAC is given by

$$R_1 + R_2 = \frac{1}{n} I(\mathbf{V}_1, \mathbf{V}_2; \mathbf{Y}') = \frac{1}{n} \left\{ h(\mathbf{Y}') - h(\mathbf{Y}'|\mathbf{V}_1, \mathbf{V}_2) \right\} \tag{35}$$

---

[1]For simplicity, we have not apply dither signals in this scheme, hence the power constraints are satisfied in the average over all the codewords, but not for each codeword. In the sequel, we use dither signal which provides the power constraint for each codeword.

$$= \frac{1}{2} \log_2 \left( \frac{\min\{P_1, P_2\}}{G(\Lambda)} \right) - \frac{1}{n} h(\mathbf{Z} \bmod \Lambda) \tag{36}$$

$$\geq \frac{1}{2} \log_2 \left( \frac{\min\{P_1, P_2\}}{G(\Lambda)} \right) - \frac{1}{n} h(\mathbf{Z}) \tag{37}$$

$$= \frac{1}{2} \log_2 \left( \frac{\min\{P_1, P_2\}}{G(\Lambda)} \right) - \frac{1}{2} \log_2 \left( 2\pi e N \right) \tag{38}$$

$$= \frac{1}{2} \log_2 \left( \frac{\min\{P_1, P_2\}}{N} \right) - \frac{1}{2} \log_2 \left( 2\pi e G(\Lambda) \right), \tag{39}$$

where (36) follows from (28) and since $\mathbf{Y}' \sim U(\mathcal{V})$; (37) follows since modulo operation decreases entropy. The theorem follows by using a lattice which is good for quantization, i.e., $G(\Lambda) \to 1/2\pi e$ as $n \to \infty$. □

The reason that lattice strategies can achieve positive rates is due to the linear structure of lattices, that fits linear additive model of the dirty MAC. This matching enables the decentralized encoders to have effectively (at the receiver) the same equivalent additive modulo channel like in the single user problem with full side information $S_1 + S_2$ at the transmitter.

### B. Lattice Strategies for General SNR

In this section we generalize the transmission scheme of Section V-A for any SNR. In the following theorem, we provide conditions under which lattice strategies are optimal.

**Theorem 3.** *Suppose that* $N \leq \sqrt{P_1 P_2} - \min\{P_1, P_2\}$ *for* $P_1 \neq P_2$. *The capacity region of the doubly dirty MAC* (6) *in the limit of strong interferences meets the outer bound of Corollary 2, and is given by the set of all rate pairs* $(R_1, R_2)$ *satisfying*

$$R_1 + R_2 \leq \frac{1}{2} \log_2 \left( 1 + \frac{\min\{P_1, P_2\}}{N} \right).$$

*Proof:* The converse part is proved in Corollary 2. The direct part involves using the canonical transmission scheme of Section IV-B with appropriate MMSE factors and dithers, the proof is given in Appendix I. □

We now consider the case that $N > \sqrt{P_1 P_2} - \min\{P_1, P_2\}$, and derive an inner bound for this case. For simplicity, we first consider the symmetric case, i.e., $P_1 = P_2 = P$ for any $N$. Using the canonical transmission scheme of Section IV-B with $\alpha_1 = \alpha_2 = \alpha_r = \alpha$, $\beta = \gamma = 1$ and $\Lambda_1 = \Lambda_2 = \Lambda_r = \Lambda$, the encoders send

$$\mathbf{X}_1 = [\mathbf{V}_1 - \alpha \mathbf{S}_1 + \mathbf{D}_1] \bmod \Lambda \tag{40}$$

$$\mathbf{X}_2 = [\mathbf{V}_2 - \alpha \mathbf{S}_2 + \mathbf{D}_2] \bmod \Lambda, \tag{41}$$

where $\mathbf{V}_1, \mathbf{V}_2 \sim \text{Unif}(\mathcal{V})$ are independent and carry the information of user 1 and user 2, respectively. Since $\mathbf{D}_1, \mathbf{D}_2 \sim \text{Unif}(\mathcal{V})$ are independent dither signals, from the dither property $\mathbf{X}_1, \mathbf{X}_2 \sim \text{Unif}(\mathcal{V})$, and hence the power constraints are satisfied. In this case the decoder is given by

$$\mathbf{Y}' = [\alpha \mathbf{Y} - \mathbf{D}_1 - \mathbf{D}_2] \bmod \Lambda. \tag{42}$$

The equivalent $\bmod - \Lambda$ MAC is given in the following lemma.

**Lemma 1 (The equivalent mod $\Lambda$ MAC).** *The equivalent channel using the encoders* (40) *and* (41) *and the decoder* (42) *is given by*

$$\mathbf{Y}' = \Big[\mathbf{V}_1 + \mathbf{V}_2 + \mathbf{Z}_{eq}\Big] \, mod \, \Lambda,\tag{43}$$

*where*

$$\mathbf{Z}_{eq} = \Big[-(1-\alpha)\mathbf{X}_1 - (1-\alpha)\mathbf{X}_2 + \alpha\mathbf{Z}\Big] \, mod \, \Lambda,\tag{44}$$

*and $\mathbf{Z}_{eq}$ is independent of $\mathbf{V_1}$ and $\mathbf{V_2}$, where $\mathbf{X}_1$, $\mathbf{X}_2$ are the self noises which are mutual independent, and independent of $\mathbf{Z}, \mathbf{V_1}, \mathbf{V_2}$*

*Proof:* The equivalent channel is given by

$$\mathbf{Y}' = \Big[\alpha(\mathbf{X}_1 + \mathbf{S}_1 + \mathbf{X}_2 + \mathbf{S}_2 + \mathbf{Z}) - \mathbf{D}_1 - \mathbf{D}_2\Big] \, mod \, \Lambda\tag{45}$$

$$= \Big[\mathbf{V}_1 + \mathbf{V}_2 - (1-\alpha)\mathbf{X}_1 - (1-\alpha)\mathbf{X}_2 + \alpha\mathbf{Z}\Big] \, mod \, \Lambda,,\tag{46}$$

where (45) follows since $Y = X_1 + S_1 + X_2 + S_2 + Z$; (46) follows from (40) and (41). Due to the dithers, the vectors $\mathbf{V_1}$, $\mathbf{V_2}$, $\mathbf{X_1}$, $\mathbf{X_2}$ are independent, and also independent of $\mathbf{Z}$. Therefore, $\mathbf{Z}_{eq}$ is independent of $\mathbf{V_1}$ and $\mathbf{V_2}$. □

From the modulo $\Lambda$ equivalent channel (43), the achievable rate sum is given by

$$R_1 + R_2 = \frac{1}{n}I(\mathbf{V}_1, \mathbf{V}_2; \mathbf{Y}')\tag{47}$$

$$= \frac{1}{n}\left\{h(\mathbf{Y}') - h(\mathbf{Y}'|\mathbf{V}_1, \mathbf{V}_2)\right\}\tag{48}$$

$$\geq \left[\frac{1}{2}\log_2\left(\frac{P}{G(\Lambda)}\right) - \frac{1}{2}\log_2\left(2\pi e(\alpha^2 N + 2(1-\alpha)^2 P)\right)\right]^+\tag{49}$$

$$= \left[\frac{1}{2}\log_2\left(\frac{P}{\alpha^2 N + 2(1-\alpha)^2 P}\right) - \frac{1}{2}\log_2\left(2\pi e G(\Lambda)\right)\right]^+\tag{50}$$

where (49) follows since a Gaussian distribution has maximal entropy for fixed variance. For the optimal $\alpha$, i.e., $\alpha^{opt} = \frac{2P}{2P+N}$, and using lattice which is good for quantization $G(\Lambda) \to 1/2\pi e$ as $n \to \infty$, we get that any rate pair

$$R_1 + R_2 \leq \left[\frac{1}{2}\log_2\left(\frac{1}{2} + \frac{P}{N}\right)\right]^+$$

is achievable, where $[x]^+ \triangleq \max\{x, 0\}$. Clearly, using time sharing argument we can achieve the following rates

$$R_1 + R_2 \leq u.c.e\left\{\left[\frac{1}{2}\log_2\left(\frac{1}{2} + \frac{P}{N}\right)\right]^+\right\},\tag{51}$$

where $u.c.e$ is the upper convex envelope with respect to $P/N$. The loss of the "half" inside the log function (instead of one) is due to the *two* independent self noises $\mathbf{X}_1$ and $\mathbf{X}_2$ that we have in the equivalent channel model as shown in Lemma 1. Nonetheless, this technique is asymptotically optimal at high SNR, when $\log(A + P/N) \approx \log(P/N)$.

For one dimensional lattice strategies, i.e., $G(\Lambda) = 1/12$, this scheme is identical to using a random binning technique with the auxiliary random variables which proposed in (34). From (50), the achievable rate sum in this case is bounded from below by

$$R_1 + R_2 \geq u.c.e \left\{ \left[ \frac{1}{2} \log_2 \left( \frac{1}{2} + \frac{P}{N} \right) - \frac{1}{2} \log_2 \left( \frac{\pi e}{6} \right) \right]^+ \right\}. \tag{52}$$

At low SNR, i.e., $\text{SNR} \leq 1/2$ ($-3\text{dB}$) the *pure* lattice strategies can not achieve any positive rates as shown in Fig. 8. Hence, timesharing is required between the point $\text{SNR} = 0$ and $\text{SNR}^*$, which is a solution of the following equation

$$\frac{df(\text{SNR})}{dSNR} = \frac{f(\text{SNR})}{\text{SNR}},$$

where $f(x) = \frac{1}{2} \log_2(0.5 + x)$. Numerical evaluation gives that $\text{SNR}^* \approx 1.655$. At low SNR, i.e., $\text{SNR} \to 0$ the inner bound is given by $R_1 + R_2 \simeq 0.425 \cdot P/N$, while the outer bound is given by $R_1 + R_2 \approx 0.721 \cdot P/N$, hence the gap between the outer bound and the inner bound is bounded by approximately 2.3 dB.

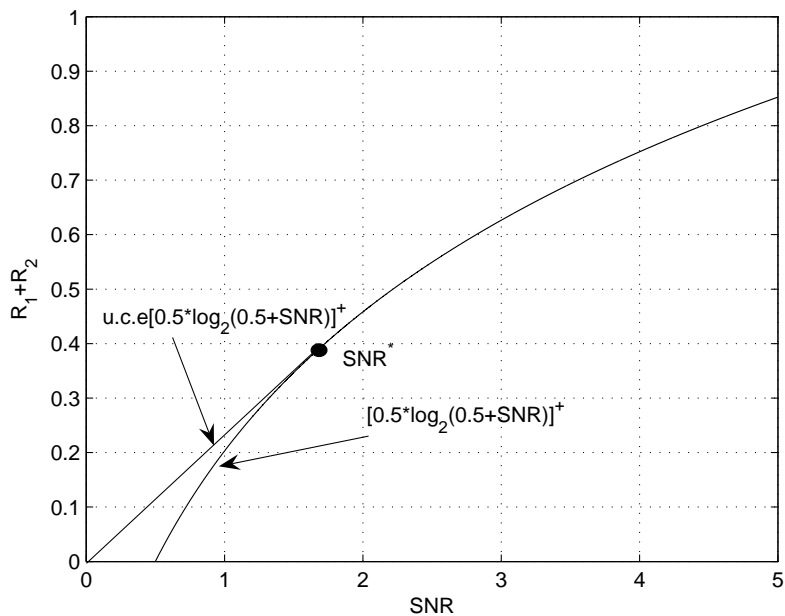

Fig. 8: Achievable rate-sum for $P_1 = P_2$.

We now derive an inner bound for the case that $N > \sqrt{P_1 P_2} - \min\{P_1, P_2\}$.

**Theorem 4.** *Suppose that $N \geq \sqrt{P_1 P_2} - \min\{P_1, P_2\}$. An achievable region for the doubly dirty MAC (6) is given for any interferences by the set of rate pairs $(R_1, R_2)$ satisfying*

$$R_1 + R_2 \leq u.c.e \left\{ \left[ \frac{1}{2} \log_2 \left( \frac{P_1 + P_2 + N}{2N + (\sqrt{P_1} - \sqrt{P_2})^2} \right) \right]^+ \right\}, \tag{53}$$

*where the upper convex envelope is with respect to $P_1$ and $P_2$.*

*Proof:* The proof is given in Appendix II □

For $N = \sqrt{P_1 P_2} - \min\{P_1, P_2\}$ the expression in (53) coincides with that in Theorem 3. While for the symmetric case $P_1 = P_2$, the region coincides with that in (51).

### C. The Gap between the Inner Bound and the Outer Bound

For $N > \sqrt{P_1 P_2} - \min\{P_1, P_2\}$, we define the gap between the outer bound (18) and the inner bound (53) as

$$\zeta(P_1, P_2) \triangleq \frac{1}{2} \log_2 \left( 1 + \frac{\min\{P_1, P_2\}}{N} \right) - u.c.e \left\{ \left[ \frac{1}{2} \log_2 \left( \frac{P_1 + P_2 + N}{2N + (\sqrt{P_1} - \sqrt{P_2})^2} \right) \right]^+ \right\}. \tag{54}$$

The gap is upper bounded by the $P_1 = P_2$ case, i.e.,

$$\zeta(P_1, P_2) \leq \zeta(P_{\min}, P_{\min}). \tag{55}$$

where $P_{\min} = \min\{P_1, P_2\}$. To see this, we fix $P_1$ and vary $P_2$ such that $P_2 \geq P_1$. The second term in the RHS of (54) is increasing in $P_2$, while the first term is constant. Therefore, we get that gap $\zeta(P_1, P_2)$ is maximized for $P_1 = P_2$. Of course for the opposite condition, i.e., $P_1 \leq P_2$, the maximum occurs again in $P_1 = P_2$.

Furthermore, by removing the upper convex envelope operation in (54), it increases the RHS of (54), hence

$$\zeta(P_1, P_2) \leq \zeta(P_{\min}, P_{\min}) \leq \max_{P_{\min}, N} \left\{ \frac{1}{2} \log_2 \left( 1 + \frac{P_{\min}}{N} \right) + \min \left\{ 0, \frac{1}{2} \log_2 \left( 1 + \frac{N - 2P_{\min}}{2P_{\min} + N} \right) \right\} \right\} \tag{56}$$

$$= \max_{P_{\min}/N} \left\{ \min \left\{ \frac{1}{2} \log_2 (1 + P_{\min}/N), \frac{1}{2} \log_2 \left( 2 \cdot \frac{1 + P_{\min}/N}{1 + 2P_{\min}/N} \right) \right\} \right\}. \tag{57}$$

Since $\log_2 (1 + P_{\min}/N)$ is increasing in $P_{\min}/N$, and $\log_2 \left( 2 \cdot \frac{1 + P_{\min}/N}{1 + 2P_{\min}/N} \right)$ is decreasing in $P_{\min}/N$, the maximum occurs when $\log_2 (1 + P_{\min}/N) = \log_2 \left( 2 \cdot \frac{1 + P_{\min}/N}{1 + 2P/N} \right)$, which is satisfied for $P_{\min}/N = 1/2$. Hence,

$$\zeta(P_1, P_2) \leq \frac{1}{2} \log_2(3/2) \approx 0.292 \text{ bit},$$

for any $P_1, P_2, N$.

In the following lemma we provide a tighter uniform upper bound for $\zeta(P_1, P_2)$.

**Lemma 2.** *Let $x^*$ be the solution of the equation $\frac{x}{x + 1/2} = \log_e(x + 1/2)$. For any $P_1, P_2, N$, the gap $\zeta(P_1, P_2)$ is bounded by*

$$\zeta(P_1, P_2) \leq \frac{\log_2 \left( \frac{1}{2} + x^* \right)}{4x^*} \approx 0.167 \text{ bit}, \tag{58}$$

*where equality holds for $P_1 = P_2 = P$, and $P/N = x^* - 0.5 \approx 1.155$.*

*Proof:* The proof is given in Appendix III □

The solution $x^*$ is evaluated numerically and it is equal to 1.655. In Fig. 9, we depict the gap for $P_1 = P_2 = P$, i.e., $\zeta(P, P)$ with respect to $P/N$.

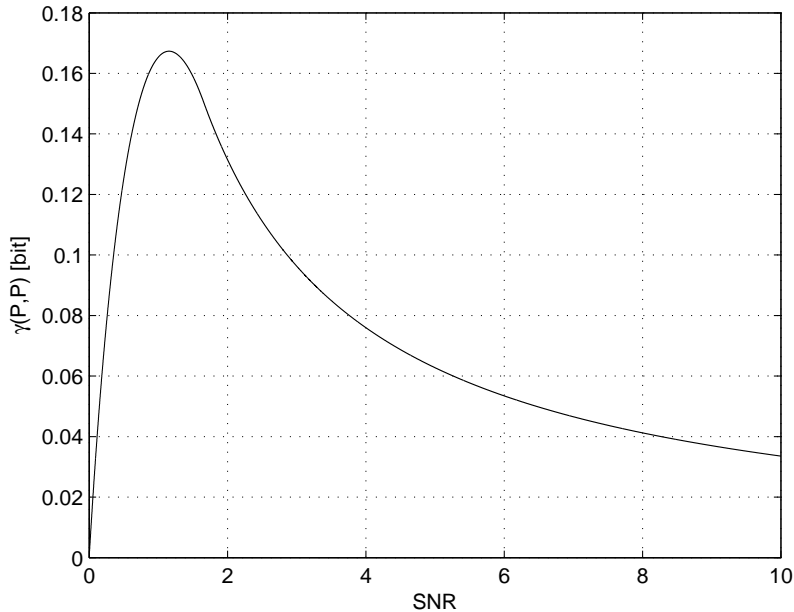

Fig. 9: The gap $\zeta(P, P)$ for $P/N \in [0, 10]$.

## VI. The Gaussian MAC with a Single Dirty User

In this section we present lattice strategies transmission scheme for the Gaussian dirty MAC with a single inform user (7) of Fig. 2. The transmission schemes used in this section are special cases of the canonical scheme of Section IV-B. Unlike in the doubly dirty MAC, the results in this section can also be achieved by random coding arguments.

In the sequel, we use the following asymptotic property for lattices which are both good for quantization (26) and for AWGN channel decoding (27).

**Lemma 3.** *Assume sequence of lattices $\Lambda_n$ with second moment $P$, which are good both for quantization* (26) *and for AWGN channel coding* (27). *Let $\mathbf{U} \sim \mathrm{Unif}(\kappa\mathcal{V})$ independent of $\mathbf{Z} \sim \mathcal{N}(0, NI_n)$, where $I_n$ is an $n \times n$ identity matrix. For any $\kappa \in [0, 1]$ which satisfies $\kappa^2 P + N = P$, we have that*

$$\lim_{n \to \infty} \frac{1}{n} h\Big([\mathbf{U} + \mathbf{Z}] \bmod \Lambda_n\Big) \geq \frac{1}{2}\log_2(2\pi eP) - \epsilon, \ \forall \ \epsilon > 0, \tag{59}$$

*Proof:* The proof is given in Appendix IV. □

### A. The Helper Problem

We first consider the *helper problem* where only user 2, the uninformed user, has a message to send and the informed user (user 1) helps user 2 to transmit at the highest possible rate, i.e., we consider the rate pair $(0, R_2)$. The upper bound for this case is given in corollary 1. In the following theorem we present the helper problem capacity for the case where $N \leq |P_1 - P_2|$.

**Theorem 5.** *Suppose that $N \leq |P_1 - P_2|$ in the MAC with a single dirty user* (7). *In the limit of strong interference, the capacity of the helper problem is given by*

$$C_{helper} = \frac{1}{2} \log_2 \left( 1 + \frac{\min\{P_1, P_2\}}{N} \right). \tag{60}$$

*Proof:* The proof is given in Appendix V. □

For $|P_1 - P_2| < N$, we derive the following inner bound.

**Lemma 4.** *Suppose that $|P_1 - P_2| < N$. An achievable rate for the* helper problem *is given by*

$$R_{helper} = u.c.e \left\{ \frac{1}{2} \log_2 \left( 1 + \frac{4 P_1 P_2}{(P_2 - P_1 + N)^2 + 4 P_1 N} \right) \right\}, \tag{61}$$

*where the upper convex envelope is with respect to $P_1$ and $P_2$. For $P_1 = P_2 = P$, this inner bound reduces to*

$$R_{helper} = u.c.e \left\{ \frac{1}{2} \log_2 \left( 1 + \text{SNR} \left( \frac{4\text{SNR}}{4\text{SNR} + 1} \right) \right) \right\}, \tag{62}$$

*where the upper convex envelope is with respect to* $\text{SNR} \triangleq \text{P/N}$.

Although the function inside the upper convex envelope operation in (61) is non-negative, by examining its Hessian matrix [17] it can be shown the this function is not convex-∩ for any $P_1$ and $P_2$ (also in (62) the function inside the upper convex envelope operation is not convex-∩ for any SNR).

*Proof:* The proof is given in Appendix VI. □

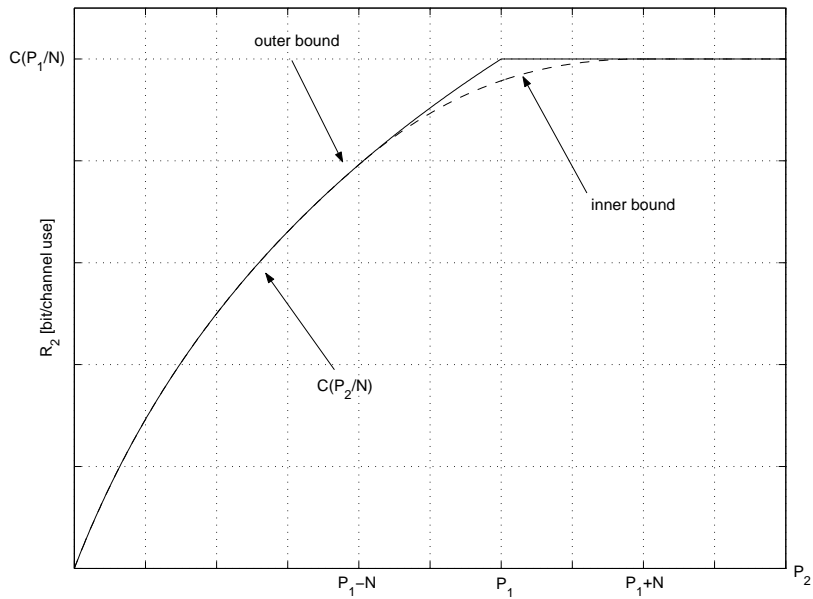

Fig. 10: Inner bound versus outer bound in the helper problem for $P_1 > N$.

In Fig. 10, the outer bound and the inner bound for the capacity of the helper problem are depicted for any $P_1, P_2, N$. As indicated in Lemma 4, there is a gap between the inner bound (61) and the outer bound (17) for $|P_1 - P_2| < N$. We define this gap as

$$\eta(P_1, P_2) \triangleq \frac{1}{2} \log_2 \left( 1 + \frac{\min\{P_1, P_2\}}{N} \right) - u.c.e \left\{ \frac{1}{2} \log_2 \left( 1 + \frac{4 P_1 P_2}{(P_2 - P_1 + N)^2 + 4 P_1 N} \right) \right\}. \tag{63}$$

In the following lemma we derive a uniform upper bound for the gap $\zeta(P_1, P_2)$.

**Lemma 5.** *For* $|P_1 - P_2| < N$*, the gap* $\eta(P_1, P_2)$ (63) *is upper bounded by*

$$\eta(P_1, P_2) \leq \eta(P_{\min}, P_{\min}) < \log_2(3) - 3/2 \approx 0.085 \ bit,$$

*where* $P_{\min} = \min\{P_1, P_2\}$*.*

*Proof:* The proof is given in Appendix VII. $\qquad\square$

We now show that at high SNR, i.e., $P_1, P_2 \gg N$ and for $|P_1 - P_2| < N$, the achievable rate $R_{\text{helper}}$ (61) meets asymptotically the outer bound (17).

**Lemma 6.** *In the limit of strong interference, the capacity of the helper problem at high SNR is given by*

$$C_{helper} = \frac{1}{2} \log_2 \left( \frac{\min\{P_1, P_2\}}{N} \right) - o(1),$$

*where* $o(1) \to 0$ *as* $P_1, P_2 \to \infty$ *for fixed* $N$*.*

*Proof:* From the achievable rate for the helper problem (61), we have that

$$R_{\text{helper}} = u.c.e \left\{ \frac{1}{2} \log_2 \left( 1 + \frac{4 P_1 P_2}{(P_2 - P_1 + N)^2 + 4 P_1 N} \right) \right\} \tag{64}$$

$$\geq \frac{1}{2} \log_2 \left( 1 + \frac{4 P_1 P_2}{4 N^2 + 4 P_1 N} \right) \tag{65}$$

$$= \frac{1}{2} \log_2 \left( 1 + \frac{\min\{P_1, P_2\}}{N} \cdot \frac{\max\{P_1, P_2\}}{P_1 + N} \right), \tag{66}$$

where (65) follows since $(P_1 - P_2 + N)^2 \leq 4 N^2$ because $|P_1 - P_2| \leq N$, and also since $\log_2 \left( 1 + \frac{4 P_1 P_2}{4 N^2 + 4 P_1 N} \right)$ is convex-$\cap$ with respect to $P_1$ and $P_2$; (66) follows since $P_1 \cdot P_2 = \min\{P_1, P_2\} \cdot \max\{P_1, P_2\}$.

Using the outer bound (17) for any $P_1, P_2, N$, the helper capacity is bounded by

$$\frac{1}{2} \log_2 \left( 1 + \frac{\min\{P_1, P_2\}}{N} \cdot \frac{\max\{P_1, P_2\}}{P_1 + N} \right) \leq C_{\text{helper}} \leq \frac{1}{2} \log_2 \left( 1 + \frac{\min\{P_1, P_2\}}{N} \right). \tag{67}$$

The lemma follows since for $P_1, P_2 \to \infty$ for fixed $N$, the LHS becomes $\frac{1}{2} \log_2 \left( 1 + \frac{\min\{P_1, P_2\}}{N} \right) - o(1)$ where $o(1) \to 0$ as $P_1, P_2 \to \infty$ for fixed $N$. $\qquad\square$

The *pure* lattice strategies are not optimal at low SNR in the helper problem, i.e. the upper convex envelope strictly increases the achievable rate in the helper problem. To see that we consider the case that $P_1 = P_2 = P$ and we show that time sharing can achieve higher rates than pure lattice strategies (the expression inside the upper convex envelope in (62)). Assume that the users coordinate their transmissions only for $1/\delta$ of the time ($\delta \geq 1$), while the rest of the time the users stay silent. During the transmission period $(1/\delta)$ user 2 transmits with power $\delta P$, while user 1 transmits half of the transmission time $(\frac{1}{2\delta})$ with power $\delta P - N$, and the rest of the time with $\delta P + N$. In this way the users satisfy the power constraints. The achievable rate of user 2 is given by

$$\begin{aligned} R_2 &= \frac{1}{2\delta} \cdot \frac{1}{2} \log_2 \left( 1 + \frac{\delta P}{N} \right) + \frac{1}{2\delta} \cdot \frac{1}{2} \log_2 \left( 1 + \frac{\delta P - N}{N} \right) \\ &= \frac{1}{4\delta} \log_2 \left( \delta \frac{P}{N} \left( 1 + \delta \frac{P}{N} \right) \right). \end{aligned}$$

Numerical evaluation shows that this expression is maximized for $\delta = 1.832\frac{N}{P}$, and the rate is given by $R_2 = 0.324 \cdot \text{SNR}$, which is higher than achievable rate using pure lattice strategies in (62) as shown in Fig. 11. This scheme is feasible only for $\text{SNR} \leq 1.832$ since $\delta \geq 1$.

For $\text{SNR} \to 0$, this inner bound goes like $O(\text{SNR})$, while the inner bound in (62) behaves like $O(\text{SNR}^2)$. On the other hand, the outer bound (17) for $\text{SNR} \to 0$ is $\lim_{\text{SNR} \to 0} \frac{1}{2}\log_2(1 + \text{SNR}) \approx 0.721 \cdot \text{SNR}$ which goes like $O(\text{SNR})$ as the inner bound.

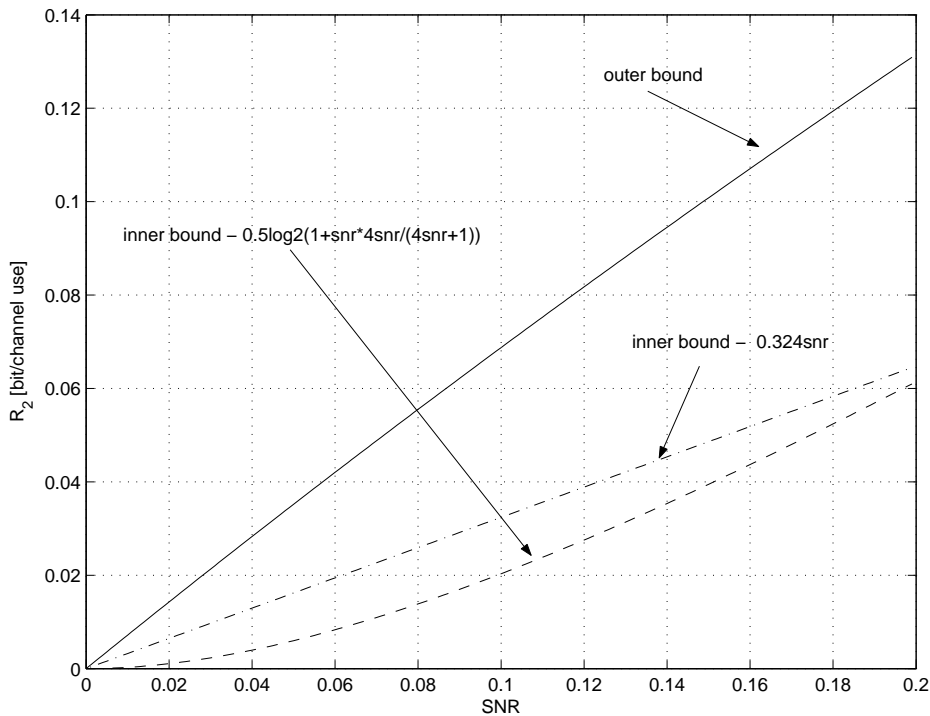

Fig. 11: Inner bounds and outer bound for helper problem at low SNR.

### B. Capacity Region at High SNR

Generally, the capacity region for the MAC with a single dirty user (7) is not known. In the following theorem we find the capacity region at high SNR, i.e., $P_1, P_2 \gg N$.

**Theorem 6.** *In the limit of strong interference, the capacity region of dirty MAC with single informed user (7) and high SNR, is given by*

$$R_2 \leq \frac{1}{2}\log_2\left(\frac{P_2}{N}\right) - o(1)$$

$$R_1 + R_2 \leq \frac{1}{2}\log_2\left(\frac{P_1}{N}\right) - o(1),$$

*where $o(1) \to 0$ as $P_1, P_2 \to \infty$.*

*Proof:* The proof is given in Appendix VIII. □

*C. Achievable Rate Region*

We now derive an achievable rate region using lattice strategies for any $P_1, P_2, N$.

**Lemma 7.** *The achievable rate region in the MAC with a single dirty user* (7) *is given by*

$$\mathcal{R} = \text{cl conv} \left\{ \bigcup_{\alpha_1 \in [0,1]} \mathcal{R}(\alpha_1) \right\}, \tag{68}$$

*and*

$$\mathcal{R}(\alpha_1) = \left\{ (R_1, R_2): \quad R_1 \leq \frac{1}{2} \log_2 \left( \frac{P1}{\min\{P_1, (1-\alpha_1)^2 P_1 + \alpha_1^2 (N + P_2)\}} \right) \right.$$
$$\left. R_2 \leq \frac{1}{2} \log_2 \left( \frac{\min\{P_1, (1-\alpha_1)^2 P_1 + \alpha_1^2 (P_2 + N)\}}{(1-\alpha_1)^2 P_1 + \alpha_1^2 N} \right) \right\} \tag{69}$$

*where* cl *and* conv *are the closure and the convex hull operations, respectively.*

*Proof:* The proof is given in Appendix IX ∎

This region is a general form which describes the achievable rate region of the MAC with a single dirty user (7), it includes the achievable rate of the helper problem, i.e., the point $(0, R_2)$ for any $P_1, P_2, N$, and also the capacity region at high SNR. This region may also be derived using random binning technique as well.

We now explore the behavior of the achievable rate region specified in Lemma 7 for several cases with respect to $P_1, P_2, N$:

**a)** For $P_1 \leq P_2 - N$: It is easily verified that the point $(R_1 = \frac{1}{2} \log_2(1 + P_1/N), 0)$ can be achieved when user 2 is silent, i.e., $X_2 = 0$ while user 1 performs point to point dirty paper coding (DPC) scheme or alternatively lattice strategies for the point to point problem as in [5]. Furthermore in Theorem 5 it was shown that for $P_1 \leq P_2 - N$, user 2 can achieve the rate $R_2 = \frac{1}{2} \log_2(1 + P_1/N)$, thus the point $(0, R_2 = \frac{1}{2} \log_2(1 + P_1/N))$ is also achievable. Therefore, time sharing between these two points achieves the outer bound (9) as shown in Fig. 12a.

**Corollary 3.** *In the limit of strong interference, for* $P_1 \leq P_2 - N$ *the capacity region of the MAC with a single dirty user* (7)*, is given by*

$$R_1 + R_2 \leq \frac{1}{2} \log_2 \left( 1 + \frac{P_1}{N} \right). \tag{70}$$

**b)** For $P_1 > P_2 - N$: This case refers to Fig. 12b-12d. We define the following rate pair

$$R_1^* \triangleq \frac{1}{2} \log_2 \left( \frac{P_1 + N}{N + \frac{P_1 P_2}{P_1 + N}} \right)$$
$$R_2^* \triangleq \frac{1}{2} \log_2 \left( 1 + \frac{P_2}{N} \cdot \frac{P_1}{P_1 + N} \right).$$

This rate pair is located on the outer bound (9) as shown in Fig. 12b-12d. To see that, it is easily verified that $R_1^* + R_2^* = \frac{1}{2} \log_2(1 + P_1/N)$ and $R_2^* < \frac{1}{2} \log_2(1 + \min\{P_1, P_2\}/N)$. On the other hand, using $\alpha_1 = \frac{P_1}{P_1 + N}$ in (69) (Lemma 7), we can achieve this rate pair. Therefore, the rate pair $(R_1^*, R_2^*)$ belongs to the boundary of the capacity region.

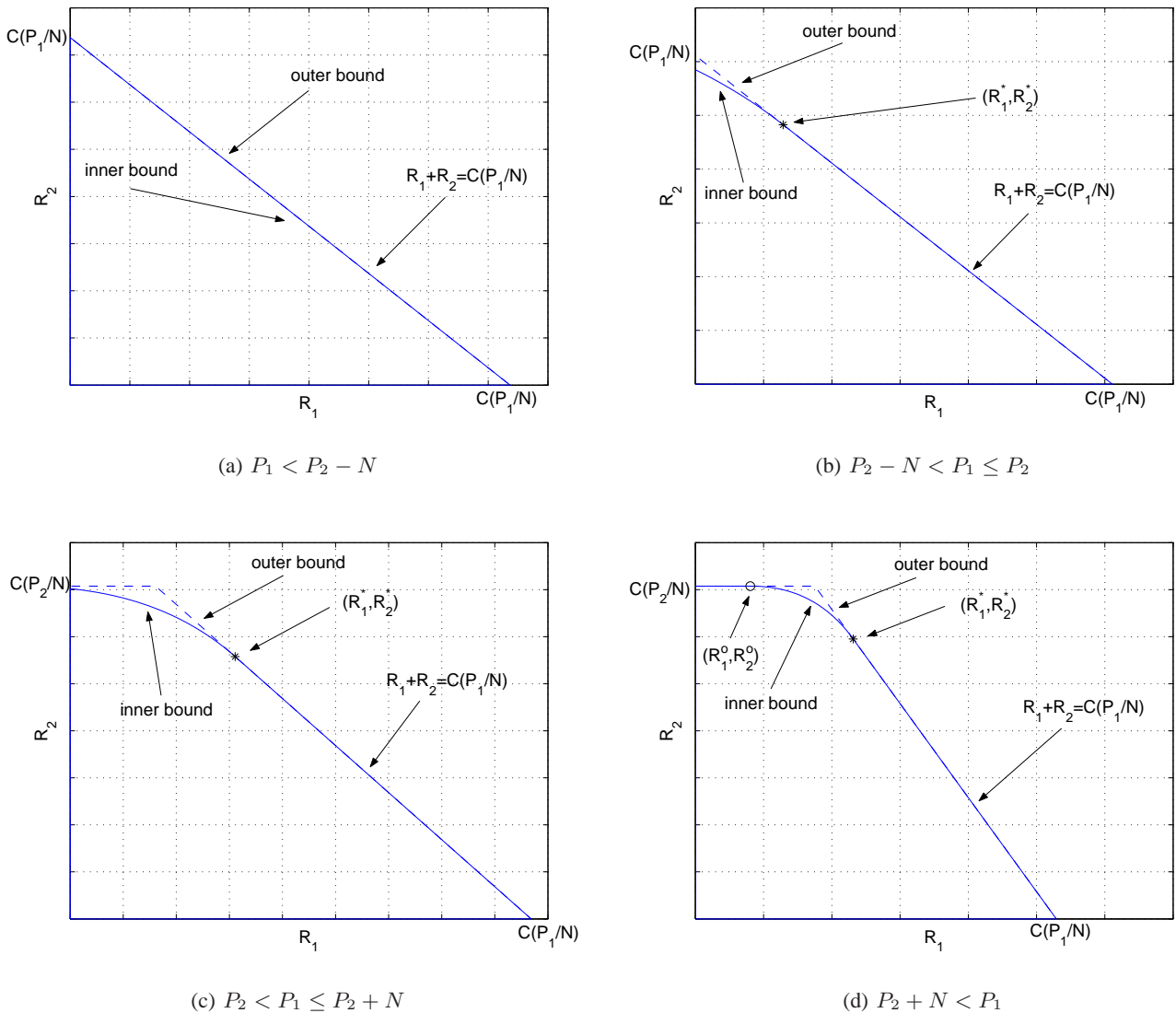

(a) $P_1 < P_2 - N$

(b) $P_2 - N < P_1 \leq P_2$

(c) $P_2 < P_1 \leq P_2 + N$

(d) $P_2 + N < P_1$

Fig. 12: Inner bound versus outer bound in the MAC with a single dirty user.

**Corollary 4.** *In the limit of strong interference, and for $P_1 > P_2 - N$, the rate pair $(R_1^*, R_2^*)$ belongs to the boundary of the capacity region in MAC with a single dirty user* (7).

The rate pair $(R_1^*, R_2^*)$ corresponds the vertex point where the inner bound and the outer bound depart from each other as shown in Fig. 12b-12d. The behavior of the achievable region versus the outer bound is shown in Fig. 12b for $P_2 - N < P_1 \leq P_2$. In this case, the gap between the inner bound and the outer bound is maximal for the helper problem, i.e., the point $(0, R_2)$, which is bounded by $\log_2(3) - 3/2 \approx 0.085$ bit (Lemma 5). In Fig. 12c, the inner bound and the outer bound for $P_2 < P_1 \leq P_2 + N$ are depicted.

**c)** For $P_2 + N < P_1$: We define the following rate pair

$$R_1^o \triangleq \frac{1}{2} \log_2 \left( \frac{P_1}{P_2 + N} \right)$$

$$R_2^o \triangleq \frac{1}{2} \log_2 \left( 1 + \frac{P_2}{N} \right).$$

Clearly, this rate pair is located on the boundary of the outer bound (9). On the other hand, using $\alpha_1 = 1$ in (69) (Lemma 7), we can achieve this rate pair as shown in Fig. 12d. In fact, it is the maximal achievable rate that user 1 can transmit while user 2 sends in its highest rate $R_2 = 0.5 \cdot \log_2(1 + P_2/N)$.

**Corollary 5.** *In the limit of strong interference, and for $P_2 + N < P_1$ the rate pair $(R_1^o, R_2^o)$ belongs to the boundary of the capacity region in MAC with a single dirty user* (7).

## VII. The Gaussian MAC with Common Interference

In this section we consider the channel in (6) where $S_1 = S_2 = S_c$, i.e., the channel state $S_c$ is known non-causally to both users. The channel model is given by

$$Y = X_1 + X_2 + S_c + Z, \tag{71}$$

where $Z \sim \mathcal{N}(0, N)$. The power constraints are $\frac{1}{n} \sum_{i=1}^n x_{1i}^2 \leq P_i$ for $i = 1, 2$. In [4], it was shown that like in the point-to-point writing on dirty paper problem, the capacity region of the dirty MAC is the same as the interference-free Gaussian MAC (*clean MAC*), i.e, the capacity region is the pentagon region [18].

The corner point $(R_1, R_2) = (0.5 \cdot \log_2(1 + \frac{P_1}{P_2 + N}), 0.5 \cdot \log_2(1 + P_2/N))$ of the pentagon is achieved by applying twice DPC for each user [10]. As in the point-to-point case, the auxiliary random variable are set to $U_1 = X_1 + \alpha_1 S_c$ where $X_1$ and $S_1$ are independent, and $U_2 = X_2 + \alpha_2 \tilde{S}_c$ where $\tilde{S}_c = (1 - \alpha_1)S_c$, and $X_2$ and $S_2$ are independent.

**a)** Writing on dirty paper for user 1 - the channel is given by

$$Y = X_1 + S_c + Z_{eq}, \tag{72}$$

where $Z_{eq} = X_2 + Z$, thus $Z_{eq}$ is independent of $X_1$ and $S_c$. Using $\alpha_1 = \frac{P_1}{P_1 + P_2 + N}$, user 1 can achieve $R_1 = 0.5 \cdot \log_2(1 + \frac{P_1}{P_2 + N})$.

**b)** Writing on dirty paper for user 2 - the equivalent channel is given by

$$Y' = Y - U_1 = X_2 + \tilde{S}_c + Z, \tag{73}$$

where $\tilde{S}_c = (1 - \alpha_1)S_c$. Using $\alpha_2 = \frac{P_2}{P_2 + N}$ user 2 can achieve $R_2 = 0.5 \cdot \log_2(1 + P_2/N)$.

Now, we focus on how to achieve the clean MAC capacity region for (71) using lattice strategies. Specifically, we show a transmission scheme for the corner point of the pentagon $(R_1, R_2) = (0.5 \cdot \log_2(1 + \frac{P_1}{P_2 + N}), 0.5 \cdot \log_2(1 + P_2/N))$ using lattice strategies. User 1 and user 2 use the lattices $\Lambda_1$ and $\Lambda_2$ with second moments $P_1$ and $P_2$, respectively. Specifically, the encoders send

$$X_1 = [V_1 - \alpha_1 S_c + D_1] \mod \Lambda_1 \tag{74}$$

$$X_2 = [V_2 - \alpha_2 \tilde{S}_c + D_2] \mod \Lambda_2, \tag{75}$$

where $\tilde{\mathbf{S}}_c = (1 - \alpha_1)\mathbf{S}_c$. The vectors $\mathbf{V}_i \sim U(\mathcal{V}_i)$ carries the information of user $i$ for $i = 1, 2$. The dither signals $\mathbf{D}_1$ and $\mathbf{D}_2$ are independent, where $\mathbf{D}_1 \sim U(\mathcal{V}_1)$ is known at the encoder of user 1 and to the decoder, and $\mathbf{D}_2 \sim U(\mathcal{V}_2)$ is known at the encoder of user 2 and to the decoder as well. From the dither quantization property the power constraints are satisfied.

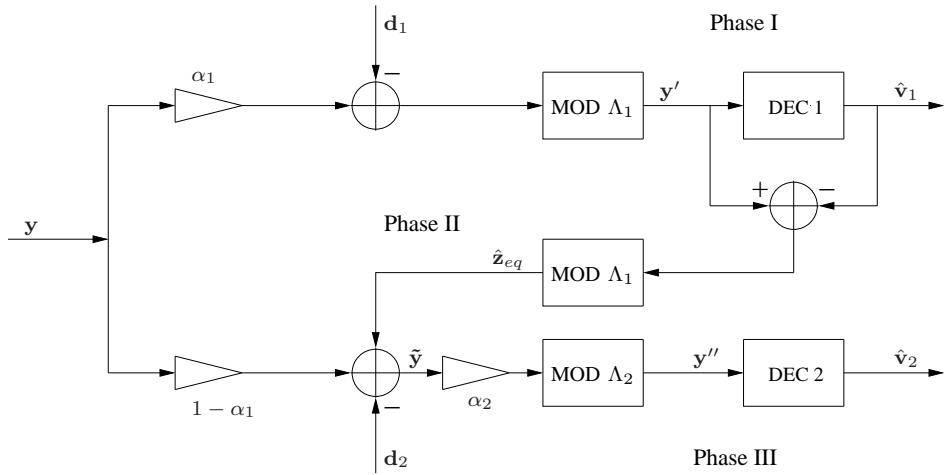

Fig. 13: Receiver for MAC with common interference.

The information bearing signals, $\mathbf{V}_1$ and $\mathbf{V}_2$, are reconstructed using a three-stage decoder as shown in Fig. 13:

**a)** *Stage I*: The decoder calculates $\mathbf{Y}' = [\alpha_1(\mathbf{Y} - \mathbf{D}_1)] \bmod \Lambda_1$. The equivalent channel is given by

$$\mathbf{Y}' = \Big[\alpha_1(\mathbf{X}_1 + \mathbf{X}_2 + \mathbf{S}_c + \mathbf{Z} - \mathbf{D}_1)\Big] \bmod \Lambda_1$$
$$= \Big[\mathbf{V}_1 - (1 - \alpha_1)\mathbf{X}_1 + \alpha_1(\mathbf{X}_2 + \mathbf{Z})\Big] \bmod \Lambda_1.$$

From the dither quantization property, $\mathbf{V}_1$ and $\mathbf{X}_1$ are independent. The rate achieved by user 1 is given by

$$R_1 = \frac{1}{n} I(\mathbf{V}_1; \mathbf{Y}') = \frac{1}{n} \left\{ h(\mathbf{Y}') - h(\mathbf{Y}'|\mathbf{V}_1)) \right\}$$
$$= \frac{1}{n} \left\{ h(\mathbf{Y}') - h([(1 - \alpha_1)\mathbf{X}_1 + \alpha_1(\mathbf{X}_2 + \mathbf{Z})] \bmod \Lambda_1) \right\}$$
$$\geq \frac{1}{2} \log_2 \left( \frac{P_1}{G(\Lambda_1)} \right) - \frac{1}{2} \log_2 \left( 2\pi e \left( (1 - \alpha_1)^2 P_1 + \alpha_1^2 (P_2 + N) \right) \right).$$

Using $\alpha_1 = \frac{P_1}{P_1 + P_2 + N}$ and lattices which are good for quantization, i.e., $G(\Lambda_1) \to 1/2\pi e$ as $n \to \infty$, any rate $R_1$ such that

$$R_1 \leq \frac{1}{2} \log_2 \left( 1 + \frac{P_1}{P_2 + N} \right) \tag{76}$$

is achievable. As a consequence, the decoder can reconstruct $\mathbf{V}_1$ with high probability.

**b)** *Stage II*: The decoder reconstructs the effective noise, i.e.,

$$\hat{\mathbf{Z}}_{eq} = [\mathbf{Y}' - \hat{\mathbf{V}}_1] \bmod \Lambda_1$$
$$= \Big[ -(1 - \alpha_1)\mathbf{X}_1 + \alpha_1(\mathbf{X}_2 + \mathbf{Z}) \Big] \bmod \Lambda_1.$$

Furthermore, with high probability we have that $\hat{\mathbf{Z}}_{eq} = -(1-\alpha_1)\mathbf{X}_1 + \alpha_1(\mathbf{X}_2 + \mathbf{Z})$, since $\frac{1}{n}E\{||-(1-\alpha_1)\mathbf{X}_1 + \alpha_1(\mathbf{X}_2 + \mathbf{Z})||^2\} = \frac{P_1(P_2+N)}{P_1+P_2+N} < P_1$.

The decoder now calculates $\mathbf{Y}_1 = \mathbf{Y} + \beta\hat{\mathbf{Z}}_{eq}$, thus

$$\mathbf{Y}_1 = \mathbf{X}_1 + \mathbf{X}_2 + \mathbf{S}_c + \mathbf{Z} - \beta(1-\alpha_1)\mathbf{X}_1 + \beta\alpha_1(\mathbf{X}_2 + \mathbf{Z})$$
$$= (1-\beta(1-\alpha_1))\mathbf{X}_1 + (1+\beta\alpha_1)\mathbf{X}_2 + \mathbf{S}_c + \mathbf{Z}(1+\beta\alpha_1).$$

For $\beta = \frac{1}{1-\alpha_1}$, we have that

$$\mathbf{Y}_1 = \frac{1}{1-\alpha_1}\mathbf{X}_2 + \mathbf{S}_c + \frac{1}{1-\alpha_1}\mathbf{Z}.$$

The receiver calculates $\tilde{\mathbf{Y}} = (1-\alpha_1)\mathbf{Y}_1$, and hence

$$\tilde{\mathbf{Y}} = \mathbf{X}_2 + \tilde{\mathbf{S}}_c + \mathbf{Z},$$

where $\tilde{\mathbf{S}}_c = (1-\alpha_1)\mathbf{S}_c$.

c) *Stage III*: The decoder calculates $\mathbf{Y}'' = [\alpha_2(\tilde{\mathbf{Y}} - \mathbf{D}_2)] \bmod \Lambda_2$. The equivalent channel is given by

$$\mathbf{Y}'' = \left[\alpha_2(\mathbf{X}_2 + \tilde{\mathbf{S}}_c + \mathbf{Z} - \mathbf{D}_2)\right] \bmod \Lambda_2$$
$$= \left[\mathbf{V}_2 - (1-\alpha_2)\mathbf{X}_2 + \alpha_1\mathbf{Z}\right] \bmod \Lambda_2.$$

Again $\mathbf{V}_2$ and $\mathbf{X}_2$ are independent. The rate achieved by user 2 is given by

$$R_2 = \frac{1}{n}I(\mathbf{V}_2; \mathbf{Y}'') = \frac{1}{n}\left\{h(\mathbf{Y}'') - h(\mathbf{Y}''|\mathbf{V}_2))\right\}$$
$$= \frac{1}{n}\left\{h(\mathbf{Y}'') - h([(1-\alpha_2)\mathbf{X}_2 + \alpha_2\mathbf{Z}] \bmod \Lambda_2)\right\}$$
$$\geq \frac{1}{2}\log_2\left(\frac{P_2}{G(\Lambda_2)}\right) - \frac{1}{2}\log_2\left(2\pi e\left((1-\alpha_1)^2 P_2 + \alpha_2^2 N\right)\right).$$

Using $\alpha_2 = \frac{P_2}{P_2+N}$ and a lattices which are good for quantization, any rate $R_2$ such that

$$R_2 \leq \frac{1}{2}\log_2\left(1 + \frac{P_2}{N}\right) \tag{77}$$

is achievable.

From symmetry, the achievability of the second corner point $(0.5 \cdot \log_2(1 + P_1/N), 0.5 \cdot \log_2(1 + \frac{P_2}{P_1+N}))$ is achieved by first decoding user 2 and then decoding user 1. The capacity region follows by using time sharing of these corner points.

## VIII. EXTENSIONS

### A. Strong Correlated Interferences

In this section we consider a generalized case for the doubly dirty MAC (6), where the interference signals are correlated. Specifically, we consider that

$$Y = X_1 + X_2 + \tilde{S}_1 + \tilde{S}_2 + Z, \tag{78}$$

where $\tilde{S}_1$ and $\tilde{S}_2$ are interference signals with a joint Gaussian distribution, i.e.,

$$\begin{pmatrix} \tilde{S}_1 \\ \tilde{S}_2 \end{pmatrix} \sim \mathcal{N} \left( \mathbf{0}, \begin{pmatrix} \tilde{\sigma}_{s_1}^2 & \rho \tilde{\sigma}_{s_1} \tilde{\sigma}_{s_2} \\ \rho \tilde{\sigma}_{s_1} \tilde{\sigma}_{s_2} & \tilde{\sigma}_{s_2}^2 \end{pmatrix} \right) \tag{79}$$

where $|\rho| < 1$ is the correlation coefficient, and $\tilde{\sigma}_{s_1}^2$ and $\tilde{\sigma}_{s_1}^2$ are the variances of $\tilde{S}_1$ and $\tilde{S}_2$, respectively. For any $\tilde{\sigma}_{s_1}, \tilde{\sigma}_{s_2}, \rho$, we denote the capacity region of (78) by $\mathcal{C}_{COR}(\tilde{\sigma}_{s_1}, \tilde{\sigma}_{s_2}, \rho)$. We also define the capacity region of the doubly dirty MAC (6) with independent Gaussian interferences $S_1$ and $S_2$ by $\mathcal{C}_{DMAC}(\sigma_{s_1}, \sigma_{s_2})$. Clearly, we have that $\mathcal{C}_{DMAC}(\sigma_{s_1}, \sigma_{s_2}) \equiv \mathcal{C}_{COR}(\sigma_{s_1}, \sigma_{s_2}, 0)$.

Generally, any joint Gaussian variables can be decomposed as

$$\tilde{S}_1 = S_1 + \beta_1 S_0 \tag{80}$$

$$\tilde{S}_2 = S_2 + \beta_2 S_0 \tag{81}$$

where $S_0 \sim \mathcal{N}(0, \sigma_{s_0}^2)$, $S_1 \sim \mathcal{N}(0, \sigma_{s_1}^2)$ and $S_2 \sim \mathcal{N}(0, \sigma_{s_2}^2)$ are independent Gaussian variables, and $\beta_1 = \text{sign}(\rho) \sqrt{|\rho|}$, $\beta_2 = \frac{\tilde{\sigma}_{s_2}}{\tilde{\sigma}_{s_1}} \sqrt{|\rho|}$ and $\sigma_{s_0}^2 = \tilde{\sigma}_{s_1}^2$. In this case we have that

$$\sigma_{s_1}^2 = \tilde{\sigma}_{s_1}^2 (1 - |\rho|) \tag{82}$$

$$\sigma_{s_2}^2 = \tilde{\sigma}_{s_2}^2 (1 - |\rho|). \tag{83}$$

The channel output can be expressed as

$$Y = X_1 + X_2 + S_1 + \beta_1 S_0 + S_2 + \beta_2 S_0 + Z \tag{84}$$

$$= X_1 + X_2 + S_1 + S_2 + S_c + Z, \tag{85}$$

where we define $S_c \triangleq (\beta_1 + \beta_2) S_0$, hence $S_1$, $S_2$, $S_c$ are Gaussian independent random variables. We denote $\mathcal{C}_{COM}(\sigma_{s_1}, \sigma_{s_2})$ to be the capacity region for the case that $(S_1, S_c)$ are known non-causally at encoder 1, and $(S_2, S_c)$ are known non-causally at encoder 2.

**Lemma 8.** *For $|\rho| < 1$, in the limit of $\tilde{\sigma}_{s_1}, \tilde{\sigma}_{s_2} \to \infty$, we have that*

$$\mathcal{C}_{COR}(\tilde{\sigma}_{s_1}, \tilde{\sigma}_{s_2}, \rho) = \mathcal{C}_{COM}(\sigma_{s_1}, \sigma_{s_2}) = \mathcal{C}_{DMAC}(\sigma_{s_1}, \sigma_{s_2}), \tag{86}$$

*where $\sigma_{s_i}^2 = \tilde{\sigma}_{s_i}^2 (1 - |\rho|)$ for $i = 1, 2$.*

*Proof:* For any $\tilde{\sigma}_{s_1}^2, \tilde{\sigma}_{s_2}^2$, we have that

$$\mathcal{C}_{DMAC}(\tilde{\sigma}_{s_1}, \tilde{\sigma}_{s_2}) = \mathcal{C}_{COR}(\tilde{\sigma}_{s_1}, \tilde{\sigma}_{s_2}, 0) \tag{87}$$

$$\subseteq \mathcal{C}_{COR}(\tilde{\sigma}_{s_1}, \tilde{\sigma}_{s_2}, \rho) \tag{88}$$

$$\subseteq \mathcal{C}_{COM}(\sigma_{s_1}, \sigma_{s_2}) \tag{89}$$

$$\subseteq \mathcal{C}_{DMAC}(\sigma_{s_1}, \sigma_{s_2}), \tag{90}$$

where (88) follows since correlation between the interferences can only increase the capacity region; (89) follows since for $S_c$ which is known to both encoders increases the capacity region; (90) follows since the capacity region increases for $S_c = 0$. The proof follows since for $\tilde{\sigma}_{s_1}^2, \tilde{\sigma}_{s_2}^2 \to \infty$, also $\sigma_{s_1}^2, \sigma_{s_2}^2 \to \infty$, hence $\mathcal{C}_{DMAC}(\sigma_{s_1}, \sigma_{s_2}) = \mathcal{C}_{DMAC}(\tilde{\sigma}_{s_1}, \tilde{\sigma}_{s_2})$. $\square$

Lemma 8 implies that for jointly Gaussian $\tilde{S}_1$ and $\tilde{S}_2$ with $|\rho| < 1$ where $\tilde{\sigma}_{s_1}, \tilde{\sigma}_{s_2} \to \infty$, the capacity region is independent of the correlation between the interferences ($\rho$). Therefore, the channel model in (78) is equivalent to the "standard" doubly dirty MAC (6) with uncorrelated $S_1$ and $S_2$. Furthermore from Lemma 8, the case that we have in addition to $S_1$ and $S_2$ a common interference $S_c$ which is known non-causally for both encoders as shown in Fig. 14, is also equivalent to doubly dirty MAC (6) in the limit of strong interferences $S_1$ and $S_2$.

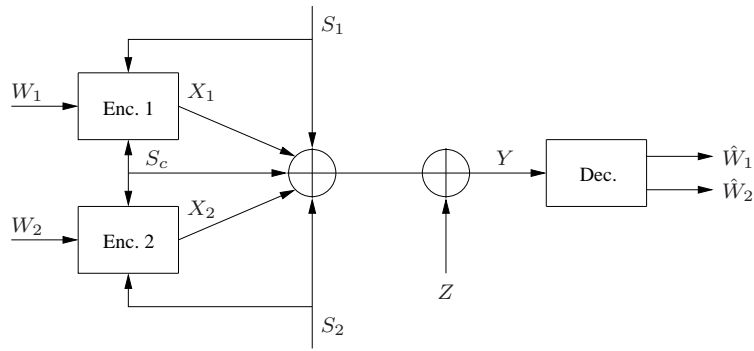

Fig. 14: MAC with private and common interferences.

### B. $K$-User Case

The results in Section V can be extended to the $K$-user case. For simplicity, we consider only the symmetric case, i.e., all the users have equal power constraints. The channel model is given by

$$Y = \sum_{i=1}^{K} X_i + \sum_{i=1}^{K} S_i + Z, \tag{91}$$

where $Z \sim \mathcal{N}(0, N)$, and the power constraint for each user is $P$. The interferences $S_i$ are independent and known non-causally only to the encoder of user $i$. Since the derivation is a straightforward extension of the two-user case we only state the final results.

**Corollary 6.** *In the limit of strong interferences the capacity region of* (91) *is contained in the following region:*

$$\sum_{i=1}^{K} R_i \leq \frac{1}{2} \log_2 \left( 1 + \frac{P}{N} \right).$$

*An achievable region for* (91) *is given by the set of all the rates satisfying*

$$\sum_{i=1}^{K} R_i \leq u.c.e \left[ \frac{1}{2} \log_2 \left( \frac{1}{K} + \frac{P}{N} \right) \right]^+.$$

Like in the two-user case Lemma 1, the factor of $1/K$ inside the log function results from the $K$ independent self noises that we have in this case. As a consequence, the rate loss between the outer bound and the inner bound increases with respect to $K$, yet the rate loss is bounded by $1/2$ bit for any $K$.

## IX. Summary

In this work we studied the Gaussian doubly dirty MAC, where each interference is known to a different transmitter. We derived an outer bound for the capacity region and found sufficient conditions under which lattice strategies meet the outer bound.

The *additive* doubly dirty MAC is a special case of channels with distributed knowledge of the channel state information among several transmitters. One may expect that in general for distributed channel state information problems, the rate loss with respect to full knowledge of the channel state at the receiver would be limited rate loss like in the (Gaussian) additive doubly dirty MAC. Generally it is not the case, to see that consider the *additive-multiplicative* model: $Y = X_1 + X_2 + S_1 \cdot S_2 + Z$, where $S_1$ and $S_2$ are known to transmitters of user 1 and user 2, respectively. In this case, for strong interferences, the uncertainty at the decoder can not be solved for any set of encoders, which indicates that the capacity tends to zero. In fact, most of the channels with effectively "strong" interferences (channels states) will result in significant rate loss.

In this work, we also studied the asymmetric case, i.e., the Gaussian dirty MAC with one interference known only at one transmitter. In particular, for the helper problem we found sufficient conditions under which lattice strategies are optimal.

We also provide a lattice based transmission scheme, which achieves the capacity region of the Gaussian MAC with common interference.

## Appendix I

### Proof of Theorem 3

In view of the outer bound (18), it is sufficient to show achievability of the point

$$(R_1, R_2) = (0, 0.5 \cdot \log_2(1 + \min\{P_1, P_2\}/N)) \,,$$

where user 1 is a helper for user 2, since from symmetry the point

$$(R_1, R_2) = (0.5 \cdot \log_2(1 + \min\{P_1, P_2\}/N), 0) \,,$$

can also be achieved. Hence, the outer bound coincides with the region which achieved by time sharing between these two points.

We first consider the case that $P_1 \geq P_2$, i.e., the helper power constraint is higher. User 1 and user 2 use lattices $\Lambda_1$ and $\Lambda_2$ with second moments $P_1$ and $P_2$, respectively. Using the canonical transmission scheme of Section IV-B with $\mathbf{V}_1 = \mathbf{0}$, $\Lambda_r = \Lambda_2$, $\alpha_1 = \beta = 1$ and $\alpha_r = \gamma = \alpha_2$, hence the encoders send

$$\mathbf{X}_1 \quad = \quad [-\mathbf{S}_1 + \mathbf{D}_1] \bmod \Lambda_1 \tag{92}$$

$$\mathbf{X}_2 \quad = \quad [\mathbf{V}_2 - \alpha_2 \mathbf{S}_2 + \mathbf{D}_2] \bmod \Lambda_2, \tag{93}$$

where $\mathbf{V}_2 \sim \text{Unif}(\mathcal{V}_2)$ carries the information of user 2; $\mathbf{D}_1$ and $\mathbf{D}_2$ are dithers signal where $\mathbf{D}_1 \sim \text{Unif}(\mathcal{V}_1)$ and $\mathbf{D}_2 \sim \text{Unif}(\mathcal{V}_2)$. User 1 mitigates the influence of the interference signal $\mathbf{S}_1$ by quantizing $\mathbf{S}_1$ with respect to the lattice $\Lambda_1 + \mathbf{d}_1$ where $\mathbf{d}_1 \in \mathbf{D}_1$. It is equivalent to use the *concentration* technique originally proposed by Willems [19].

The receiver calculates $\mathbf{Y}' = [\alpha_2(\mathbf{Y} - \mathbf{D}_1) - \mathbf{D}_2] \bmod \Lambda_2$. The equivalent channel is given by

$$\mathbf{Y}' = \left[ \alpha_2(\mathbf{X}_1 + \mathbf{S}_1 + \mathbf{X}_2 + \mathbf{s}_2 + \mathbf{Z} - \mathbf{D}_1) - \mathbf{D}_2 \right] \bmod \Lambda_2 \tag{94}$$

$$= \left[ \alpha_2[\mathbf{X}_2 + \mathbf{S}_2] - \mathbf{D}_2 - \alpha_2 Q_{\Lambda_1}(-\mathbf{S}_1 + \mathbf{D}_1) \right] \bmod \Lambda_2 \tag{95}$$

$$= \left[ \mathbf{V}_2 - (1 - \alpha_2)\mathbf{X}_2 + \alpha_2\mathbf{Z} - \alpha_2 Q_{\Lambda_1}(-\mathbf{S}_1 + \mathbf{D}_1) \right] \bmod \Lambda_2, \tag{96}$$

where (95) follows from (92); (96) follows from (93).

In order to achieve the maximal rate, the optimal MMSE factor is used, i.e., $\alpha_2 = \frac{P_2}{P_2 + N}$. For $\Lambda_2 = \alpha_2\Lambda_1$, we have that $\alpha_2 Q_{\Lambda_1}(-\mathbf{S}_1 + \mathbf{D}_1) \in \Lambda_2$. Such a selection of lattices causes the element $\alpha_2 Q_{\Lambda_1}(\mathbf{D}_1 - \mathbf{S}_1)$ to disappear after the modulo $\Lambda_2$ operation. However, it restricts the users powers to be $P_2 = \alpha_2^2 P_1$, hence the equivalent channel is given by

$$\mathbf{Y}' = \left[ \mathbf{V}_2 - (1 - \alpha_2)\mathbf{X}_2 + \alpha_2\mathbf{Z} \right] \bmod \Lambda_2, \tag{97}$$

From the dithered quantization property, $\mathbf{V}_2$ and $\mathbf{X}_2$ are independent, hence the rate achieved by user 2 is given by

$$R_2 = \frac{1}{n} I(\mathbf{V}_2; \mathbf{Y}') = \frac{1}{n} \left\{ h(\mathbf{Y}') - h(\mathbf{Y}'|\mathbf{V}_2) \right\}$$

$$= \frac{1}{n} \left\{ h(\mathbf{Y}') - h([(1 - \alpha_2)\mathbf{X}_2 + \alpha_2\mathbf{Z}] \bmod \Lambda_2) \right\}$$

$$\geq \frac{1}{2} \log_2 \left( \frac{P_2}{G(\Lambda_2)} \right) - \frac{1}{2} \log_2 \left( 2\pi e \left( (1 - \alpha_2)^2 P_2 + \alpha_2^2 N \right) \right).$$

Using lattices which are good for quantization (26), i.e., $G(\Lambda_1), G(\Lambda_2) \to 1/2\pi e$ as $n \to \infty$, and for $\alpha_2 = \frac{P_2}{P_2 + N}$, we get that any rate

$$R_2 \quad \leq \quad \frac{1}{2} \log_2 \left( 1 + \frac{P_2}{N} \right), \tag{98}$$

is achievable. Clearly, for $P_1 = P_2 \left( \frac{P_2 + N}{P_2} \right)^2$ the inner bound meets the outer bound (18). Likewise, for $P_1 \geq P_2(\frac{P_2+N}{P_2})^2$, the outer bound (18) remains $\frac{1}{2} \log_2(1 + P_2/N)$, thus the outer bound is also achievable.

We now consider the case that $P_1 < P_2$. Using the canonical transmission scheme of Section IV-B with $\mathbf{V}_1 = \mathbf{0}$, $\mathbf{D}_2 = \mathbf{0}$, $\alpha_2 = \gamma = 1$, $\Lambda_r = \Lambda_1$ and $\alpha_r = \alpha_1$, the encoders send

$$\mathbf{X}_1 = [-\alpha_1\mathbf{S}_1 + \mathbf{D}_1] \bmod \Lambda_1 \tag{99}$$

$$\mathbf{X}_2 = [\mathbf{V}_2 - \mathbf{S}_2] \bmod \Lambda_2, \tag{100}$$

where $\mathbf{V}_2 \sim \text{Unif}(\mathcal{V}_2)$ carries the information of user 2, the dither signal $\mathbf{D}_1 \sim \text{Unif}(\mathcal{V}_1)$ is known at the encoder of user 1 and to the decoder. The receiver calculates $\mathbf{Y}' = [\alpha_1\mathbf{Y} - \mathbf{D}_1] \bmod \Lambda_1$. The equivalent channel is given

by

$$\mathbf{Y}' = \left[\alpha_1(\mathbf{X}_1 + \mathbf{S}_1 + \mathbf{X}_2 + \mathbf{S}_2 + \mathbf{z}) - \mathbf{D}_1\right] \bmod \Lambda_1 \tag{101}$$

$$= \left[\alpha_1\mathbf{V}_2 + \alpha_1(\mathbf{X}_1 + \mathbf{S}_1) + \alpha_1\mathbf{Z} - \alpha_1 Q_{\Lambda_2}(\mathbf{V}_2 - \mathbf{S}_2) - \mathbf{D}_1\right] \bmod \Lambda_1 \tag{102}$$

$$= \left[\alpha_1\mathbf{V}_2 - (1 - \alpha_1)\mathbf{X}_1 + \alpha_1\mathbf{Z} - \alpha_1 Q_{\Lambda_2}(\mathbf{V}_2 - \mathbf{S}_2)\right] \bmod \Lambda_1, \tag{103}$$

where (102) follows from (100); (103) follows from (99).

For $\alpha_1 = \frac{P_1}{P_1 + N}$, and $\Lambda_1 = \alpha_1\Lambda_2$, we have that $\alpha_1 Q_{\Lambda_2}(\mathbf{V}_2 - \mathbf{S}_2) \in \Lambda_1$. Such a selection of lattices causes the element $\alpha_1 Q_{\Lambda_2}(\mathbf{V}_2 - \mathbf{S}_2)$ to disappear after the modulo $\Lambda_1$ operation. However, it restricts the user powers to be $P_1 = \left(\frac{P_1}{P_1 + N}\right)^2 P_2$. As a consequence, we have that

$$\mathbf{Y}' = [\alpha_1\mathbf{V}_2 - (1 - \alpha_1)\mathbf{X}_1 + \alpha_1\mathbf{Z}] \bmod \Lambda_1, \tag{104}$$

where $\alpha_1\mathbf{V}_2 \sim \text{Unif}(\mathcal{V}_1)$. Since $\mathbf{V}_2$ and $\mathbf{X}_1$ are independent, the rate achieved by user 2 is given by

$$R_2 = \frac{1}{n} I(\mathbf{V}_2; \mathbf{Y}') = \frac{1}{n} \left\{h(\mathbf{Y}') - h(\mathbf{Y}'|\mathbf{V}_2)\right\}$$

$$= \frac{1}{n} \left\{h(\mathbf{Y}') - h([(1 - \alpha_1)\mathbf{X}_1 + \alpha_1\mathbf{Z}] \bmod \Lambda_1)\right\}$$

$$\geq \frac{1}{2}\log_2\left(\frac{P_1}{G(\Lambda_1)}\right) - \frac{1}{2}\log_2\left(2\pi e\left((1 - \alpha_1)^2 P_1 + \alpha_1{}^2 N\right)\right) \tag{105}$$

$$= \frac{1}{2}\log_2\left(1 + \frac{P_1}{N}\right) - \frac{1}{2}\log_2\left(2\pi e G(\Lambda_1)\right). \tag{106}$$

Using lattices which are good for quantization, i.e., $G(\Lambda_1), G(\Lambda_2) \to 1/2\pi e$ as $n \to \infty$, we get that any rate

$$R_2 \leq \frac{1}{2}\log_2\left(1 + \frac{P_1}{N}\right), \tag{107}$$

is achievable. Therefore, for $P_2 = P_1(\frac{P_1 + N}{P_1})^2$ the inner bound meets the outer bound (18). For $P_2 \geq P_1(\frac{P_1 + N}{P_1})^2$, the outer bound (18) remains $\frac{1}{2}\log_2(1 + P_1/N)$, therefore the outer bound is also achievable.

From (107) and (98), the following rate is achievable for the point$(0, R_2)$:

$$R_2 = \begin{cases} \frac{1}{2}\log_2\left(1 + \frac{P_1}{N}\right), & P_2 \geq P_1\left(\frac{P_1 + N}{P_1}\right)^2 \\ \frac{1}{2}\log_2\left(1 + \frac{P_2}{N}\right), & P_1 \geq P_2\left(\frac{P_2 + N}{P_2}\right)^2 \end{cases} \tag{108}$$

Due to the symmetry between users, the same arguments can be used to show the achievable rate for the point $(R_1, 0)$:

$$R_1 = \begin{cases} \frac{1}{2}\log_2\left(1 + \frac{P_2}{N}\right), & P_1 \geq P_2\left(\frac{P_2 + N}{P_2}\right)^2 \\ \frac{1}{2}\log_2\left(1 + \frac{P_1}{N}\right), & P_2 \geq P_1\left(\frac{P_1 + N}{P_1}\right)^2 \end{cases} \tag{109}$$

The theorem follows since any rate pair in the straight line $R_1 + R_2 = \frac{1}{2}\log_2(1 + \min\{P_1, P_2\}/N)$ is achievable using time sharing between (108) and (109).

APPENDIX II

PROOF OF THEOREM 4

Clearly, it is only required to show the achievable region inside the upper convex envelope operation in (53), since the region including the upper convex envelope may achieve using time sharing.

We first consider the case that $P_1 \leq P_2$. User 1 and user 2 use the lattices $\Lambda_1$ and $\Lambda_2$ with second moments $P_1$ and $P_2$, respectively. We show achievability for the rate pair $(R_1, 0)$ where

$$R_1 = \frac{1}{2} \log_2 \left( \frac{P_1 + P_2 + N}{2N + (\sqrt{P_1} - \sqrt{P_2})^2} \right).$$

Using the canonical transmission scheme of Section IV-B with $\mathbf{V}_2 = \mathbf{0}$, $\Lambda_r = \Lambda_1$, $\gamma = 1$ and $\alpha_r = \alpha_1$, hence the encoders send

$$\mathbf{X}_1 = [\mathbf{V}_1 - \alpha_1 \mathbf{S}_1 + \mathbf{D}_1] \bmod \Lambda_1 \tag{110}$$

$$\mathbf{X}_2 = [-\alpha_2 \mathbf{S}_2 + \mathbf{D}_2] \bmod \Lambda_2, \tag{111}$$

where $\mathbf{V}_1 \sim \mathrm{Unif}(\mathcal{V}_1)$ carries the information of user 1; $\mathbf{D}_1 \sim \mathrm{Unif}(\mathcal{V}_1)$ and $\mathbf{D}_2 \sim \mathrm{Unif}(\mathcal{V}_2)$ are the dithers. Due to the dither quantization property, the power of the transmitted signals are satisfied. The receiver calculates $\mathbf{Y}' = [\alpha_1 \mathbf{Y} - \mathbf{D}_1 - \beta \mathbf{D}_2] \bmod \Lambda_1$ where $\beta = \frac{\alpha_1}{\alpha_2}$. The equivalent channel is given by

$$\mathbf{Y}' = \left[ \alpha_1 (\mathbf{X}_1 + \mathbf{S}_1 + \mathbf{X}_2 + \mathbf{S}_2 + \mathbf{Z}) - \mathbf{D}_1 - \beta \mathbf{D}_2 \right] \bmod \Lambda_1 \tag{112}$$

$$= \left[ \mathbf{V}_1 - (1 - \alpha_1)\mathbf{X}_1 + \alpha_1 \mathbf{Z} + \alpha_1(\mathbf{X}_2 + \mathbf{S}_2) - \beta \mathbf{D}_2 \right] \bmod \Lambda_1 \tag{113}$$

$$= \left[ \mathbf{V}_1 - (1 - \alpha_1)\mathbf{X}_1 + \alpha_1 \mathbf{Z} + \alpha_1(1 - \alpha_2)\mathbf{S}_2 - (\beta - \alpha_1)\mathbf{D}_2 - \alpha_1 Q_{\Lambda_2}(-\alpha_2 \mathbf{S}_2 + \mathbf{D}_2) \right] \bmod \Lambda_1 \tag{114}$$

$$= \left[ \mathbf{V}_1 - (1 - \alpha_1)\mathbf{X}_1 + \alpha_1 \mathbf{Z} - \frac{\alpha_1}{\alpha_2}(1 - \alpha_2)[-\alpha_2 \mathbf{S}_2 + \mathbf{D}_2 - Q_{\Lambda_2}(-\alpha_2 \mathbf{S}_2 + \mathbf{D}_2)] - \frac{\alpha_1}{\alpha_2} Q_{\Lambda_2}(-\alpha_2 \mathbf{s}_2 + \mathbf{d}_2) \right] \bmod \Lambda_1 \tag{115}$$

$$= \left[ \mathbf{V}_1 - (1 - \alpha_1)\mathbf{X}_1 + \alpha_1 \mathbf{Z} - \frac{\alpha_1}{\alpha_2}(1 - \alpha_2)\mathbf{X}_2 - \frac{\alpha_1}{\alpha_2} Q_{\Lambda_2}(-\alpha_2 \mathbf{S}_2 + \mathbf{D}_2) \right] \bmod \Lambda_1, \tag{116}$$

where (113) follows from (110); (114) follows from (111); (115) follows since $\beta = \frac{\alpha_1}{\alpha_2}$; (116) follows from (111). For $\Lambda_1 = \frac{\alpha_1}{\alpha_2}\Lambda_2$, hence $\frac{\alpha_1}{\alpha_2} = \sqrt{\frac{P_1}{P_2}}$ and $\frac{\alpha_1}{\alpha_2} Q_{\Lambda_2}(-\alpha_2 \mathbf{S}_2 + \mathbf{D}_2) \in \Lambda_1$. Such a selection of lattices causes the element $\frac{\alpha_1}{\alpha_2} Q_{\Lambda_2}(-\alpha_2 \mathbf{S}_2 + \mathbf{D}_2)$ to disappear after the modulo $\Lambda_1$ operation in (116). As a consequence, the equivalent channel is given by

$$\mathbf{Y}' = \left[ \mathbf{V}_1 - (1 - \alpha_1)\mathbf{X}_1 + \alpha_1 \mathbf{Z} - \left( 1 - \alpha_1 \sqrt{\frac{P_2}{P_1}} \right) \cdot \sqrt{\frac{P_1}{P_2}} \mathbf{X}_2 \right] \bmod \Lambda_1, \tag{117}$$

where $\sqrt{\frac{P_1}{P_2}} \cdot \mathbf{X}_2 \in \Lambda_1$. Using lattices which are good for quantization with $G(\Lambda_1), G(\Lambda_2) \to 1/2\pi e$ as $n \to \infty$,

we get that

$$
\begin{aligned}
R_1 = \frac{1}{n} I(\mathbf{V}_1; \mathbf{Y}') &= \frac{1}{n} \left\{ h(\mathbf{Y}') - h(\mathbf{Y}'|\mathbf{V}_1) \right\} \\
&= \frac{1}{n} \left\{ h(\mathbf{Y}') - h\left( \left[ (1-\alpha_1)\mathbf{X}_1 + \alpha_1 \mathbf{Z} - \sqrt{\frac{P_1}{P_2}} \left( 1 - \alpha_1 \sqrt{\frac{P_2}{P_1}} \right) \mathbf{X}_2 \right] \bmod \Lambda_1 \right) \right\} \\
&\geq \left[ \frac{1}{2} \log_2 \left( \frac{P_1}{(1-\alpha_1)^2 P_1 + \alpha_1^2 N + (\sqrt{P_1} - \alpha_1 \sqrt{P_2})^2} \right) \right]^+.
\end{aligned}
\tag{118}
$$

The optimal $\alpha_1$ that maximizes $R_1$ is given by $\alpha_1 = \frac{\sqrt{P_1}(\sqrt{P_1} + \sqrt{P_2})}{P_1 + P_2 + N}$, in this case we get that any rate

$$
R_1 \leq \left[ \frac{1}{2} \log_2 \left( \frac{P_1 + P_2 + N}{2N + (\sqrt{P_2} - \sqrt{P_1})^2} \right) \right]^+
\tag{119}
$$

is achievable.

We now consider the case that $P_1 \geq P_2$. Again, we show achievability for the rate pair $(R_1, 0)$ where

$$
R_1 = \frac{1}{2} \log_2 \left( \frac{P_1 + P_2 + N}{2N + (\sqrt{P_1} - \sqrt{P_2})^2} \right).
$$

Using the canonical transmission scheme of Section IV-B with $\mathbf{V}_2 = \mathbf{0}$, $\Lambda_r = \Lambda_2$, $\beta = 1$ and $\alpha_r = \alpha_2$, the encoders send

$$
\mathbf{X}_1 = [\mathbf{V}_1 - \alpha_1 \mathbf{S}_1 + \mathbf{D}_1] \bmod \Lambda_1
\tag{120}
$$

$$
\mathbf{X}_2 = [-\alpha_2 \mathbf{S}_2 + \mathbf{D}_2] \bmod \Lambda_2,
\tag{121}
$$

where $\mathbf{V}_1 \sim \mathrm{Unif}(\mathcal{V}_1)$ carries the information of user 1; $\mathbf{D}_1 \sim \mathrm{Unif}(\mathcal{V}_1)$ and $\mathbf{D}_2 \sim \mathrm{Unif}(\mathcal{V}_2)$ are the dithers. Due to the dither quantization property, the power constraints are satisfied. The receiver calculates $\mathbf{Y}' = [\alpha_2 \mathbf{y} - \mathbf{D}_2 - \gamma \mathbf{D}_1] \bmod \Lambda_2$ where $\gamma = \frac{\alpha_2}{\alpha_1}$. The equivalent channel is given by

$$
\mathbf{Y}' = \left[ \alpha_2 (\mathbf{x}_1 + \mathbf{S}_1 + \mathbf{X}_2 + \mathbf{s}_2 + \mathbf{Z}) - \mathbf{D}_2 - \gamma \mathbf{D}_1 \right] \bmod \Lambda_2
\tag{122}
$$

$$
= \left[ (1-\alpha_2)\mathbf{X}_2 + \alpha_2 \mathbf{Z} + \alpha_2 (\mathbf{X}_1 + \mathbf{s}_1) - \gamma \mathbf{D}_1 \right] \bmod \Lambda_2
\tag{123}
$$

$$
= \left[ -(1-\alpha_2)\mathbf{X}_2 + \alpha_2 \mathbf{Z} + \alpha_2 [\mathbf{V}_1 + (1-\alpha_1)\mathbf{S}_1 - Q_{\Lambda_1}(\mathbf{V}_1 - \alpha_1 \mathbf{S}_1 + \mathbf{D}_1)] - (\gamma - \alpha_2)\mathbf{D}_1 \right] \bmod \Lambda_2
\tag{124}
$$

$$
= \left[ \frac{\alpha_2}{\alpha_1} \mathbf{V}_1 - (1-\alpha_2)\mathbf{X}_2 + \alpha_2 \mathbf{Z} - \frac{\alpha_2}{\alpha_1}(1-\alpha_1)[\mathbf{V}_1 - \alpha_1 \mathbf{S}_1 + \mathbf{D}_1 - Q_{\Lambda_1}(\mathbf{V}_1 - \alpha_1 \mathbf{S}_1 + \mathbf{D}_1)] \right.
$$
$$
\left. - \frac{\alpha_2}{\alpha_1} Q_{\Lambda_1}(\mathbf{V}_1 - \alpha_1 \mathbf{S}_1 + \mathbf{D}_1) \right] \bmod \Lambda_2
\tag{125}
$$

$$
= \left[ \frac{\alpha_2}{\alpha_1} \mathbf{V}_1 - \frac{\alpha_2}{\alpha_1}(1-\alpha_1)\mathbf{X}_1 - (1-\alpha_2)\mathbf{X}_2 + \alpha_2 \mathbf{Z} - \frac{\alpha_2}{\alpha_1} Q_{\Lambda_1}(\mathbf{V}_1 - \alpha_1 \mathbf{S}_1 + \mathbf{D}_1) \right] \bmod \Lambda_2
\tag{126}
$$

where (123) follows from (121); (124) follows from (120); (125) follows since $\gamma = \frac{\alpha_2}{\alpha_1}$; (126) follows from (120). For $\Lambda_2 = \frac{\alpha_2}{\alpha_1} \Lambda_1$, hence $\frac{\alpha_1}{\alpha_2} = \sqrt{\frac{P_1}{P_2}}$ and $\frac{\alpha_2}{\alpha_1} Q_{\Lambda_1}(\mathbf{V}_1 - \alpha_1 \mathbf{S}_1 + \mathbf{D}_1) \in \Lambda_2$. Such a selection of lattices causes the element $\frac{\alpha_2}{\alpha_1} Q_{\Lambda_1}(\mathbf{V}_1 - \alpha_1 \mathbf{S}_1 + \mathbf{D}_1)$ to disappear after the modulo $\Lambda_2$ operation in (126). As a consequence, the equivalent channel is given by

$$
\mathbf{Y}' = \left[ \sqrt{\frac{P_2}{P_1}} \mathbf{V}_1 - \left( 1 - \alpha_2 \sqrt{\frac{P_1}{P_2}} \right) \cdot \sqrt{\frac{P_2}{P_1}} \mathbf{X}_1 - (1-\alpha_2)\mathbf{X}_2 + \alpha_2 \mathbf{Z} \right] \bmod \Lambda_2,
\tag{127}
$$

where $\sqrt{\frac{P_2}{P_1}}\mathbf{V}_1 \in \Lambda_2$ and $\sqrt{\frac{P_2}{P_1}}\mathbf{X}_1 \in \Lambda_2$. Using lattices which are good for quantization, i.e., $G(\Lambda_1), G(\Lambda_2) \to 1/2\pi e$ as $n \to \infty$, we have that

$$
\begin{aligned}
R_1 &= \frac{1}{n}I(\mathbf{V}_1; \mathbf{Y}') = \frac{1}{n}\left\{h(\mathbf{Y}') - h(\mathbf{Y}'|\mathbf{V}_1))\right\} \\
&= \frac{1}{n}\left\{h(\mathbf{Y}') - h\left(\left[\sqrt{\frac{P_2}{P_1}}\left(1 - \alpha_2\sqrt{\frac{P_1}{P_2}}\right)\mathbf{X}_1 - (1-\alpha_2)\mathbf{X}_2 + \alpha_2\mathbf{Z}\right] \bmod \Lambda_2\right)\right\} \\
&\geq \left[\frac{1}{2}\log_2\left(\frac{P_2}{\left(\sqrt{P_2} - \alpha_2\sqrt{P_1}\right)^2 + (1-\alpha_2)^2 P_2 + \alpha_2^2 N}\right)\right]^+.
\end{aligned}
\tag{128}
$$

The rate $R_1$ is maximized for $\alpha_2 = \frac{\sqrt{P_2}(\sqrt{P_1} + \sqrt{P_2})}{P_1 + P_2 + N}$, in this case we get that any rate

$$
R_1 \leq \left[\frac{1}{2}\log_2\left(\frac{P_1 + P_2 + N}{2N + (\sqrt{P_1} - \sqrt{P_2})^2}\right)\right]^+
\tag{129}
$$

is achievable, which is identical to the case that $P_1 \leq P_2$ (119). Therefore, the achievable rate of the point $(R_1, 0)$ for $N \geq \sqrt{P_1 P_2} - \min\{P_1, P_2\}$ is given by.

$$
(R_1, 0) = \left(\left[\frac{1}{2}\log_2\left(\frac{P_1 + P_2 + N}{2N + (\sqrt{P_1} - \sqrt{P_2})^2}\right)\right]^+, 0\right).
\tag{130}
$$

Due to the symmetry, it can be shown that the achievable rate of the point $(0, R_2)$ for $N \geq \sqrt{P_1 P_2} - \min\{P_1, P_2\}$ is given by

$$
(0, R_2) = \left(0, \left[\frac{1}{2}\log_2\left(\frac{P_1 + P_2 + N}{2N + (\sqrt{P_1} - \sqrt{P_2})^2}\right)\right]^+\right).
\tag{131}
$$

The theorem follows by using a time sharing between the achievable rate pairs in (130) and (131).

## Appendix III

## Proof of Lemma 2

Without loss of generality we can assume that $P_1 \leq P_2$ for $N > \sqrt{P_1 P_2} - \min\{P_1, P_2\}$. From (55), we have that

$$
\gamma(P_1, P_2) \leq \gamma(P_1, P_1) = \frac{1}{2}\log_2\left(1 + \frac{P_1}{N}\right) - u.c.e\left\{\left[\frac{1}{2}\log_2\left(\frac{1}{2} + \frac{P_1}{N}\right)\right]^+\right\}.
\tag{132}
$$

Let us define that $x \triangleq \frac{P_1}{N}$, thus $\gamma(P_1, P_1)$ becomes

$$
\gamma(x) = \frac{1}{2}\log_2(1 + x) - u.c.e\left\{\left[\frac{1}{2}\log_2\left(\frac{1}{2} + x\right)\right]^+\right\},
\tag{133}
$$

where the upper convex envelope is with respect to $x$. We also define the following function

$$
f(x) \triangleq \frac{1}{2}\log_2\left(\frac{1}{2} + x\right).
\tag{134}
$$

The function $[f(x)]^+$ is not a convex - $\cap$ function with respect to $x$. We define the point $x^*$, such that the upper convex envelope of $[f(x)]^+$ is achieved by time-sharing between the points $x = 0$ and $x = x^*$, therefore we have that

$$\frac{\partial f(x = x^*)}{\partial x} = \frac{\frac{1}{2}\log_2(e)}{\frac{1}{2} + x^*} = \frac{\frac{1}{2}\log_2\left(\frac{1}{2} + x^*\right)}{x^*} \tag{135}$$

Therefore,

$$u.c.e\left\{[f(x)]^+\right\} = \begin{cases} \frac{1}{2}\log_2\left(\frac{1}{2} + x\right), & x \geq x^* \\ C^* x, & 0 \leq x \leq x^* \end{cases} \tag{136}$$

where $C^* \triangleq \frac{\frac{1}{2}\log_2(e)}{\frac{1}{2} + x^*}$. The value of $x^*$ can be evaluated (numerically) from the equation $C^* x^* = \frac{1}{2}\log_2\left(\frac{1}{2} + x^*\right)$, which results that $x^* \approx 1.655$.

**a)** For $x \geq x^*$: $\gamma(x)$ is given by

$$\gamma(x) = \frac{1}{2}\log_2\left(\frac{1 + x}{\frac{1}{2} + x}\right) = \frac{1}{2}\log_2\left(1 + \frac{\frac{1}{2}}{1 + x}\right). \tag{137}$$

Since $\gamma(x)$ is decreasing with respect to $x$, hence $\gamma(x)$ is maximized for $x = x^*$.

**b)** For $0 \leq x \leq x^*$: $\gamma(x)$ is given by

$$\gamma(x) = \frac{1}{2}\log_2\left(1 + x\right) - C^* x. \tag{138}$$

The maximum of $\gamma(x)$ occurs at $x^* - \frac{1}{2}$, hence we get that

$$\gamma(x) \leq \gamma\left(x^* - \frac{1}{2}\right) = \frac{\frac{1}{2}\log_2\left(\frac{1}{2} + x^*\right)}{2x^*}. \tag{139}$$

The lemma follows since $\gamma(x^*) \leq \gamma(x^* - 1/2)$.

## Appendix IV

### Proof of Lemma 3

Let us define the following probabilities:

$$P_{\overline{\mathcal{V}}} \triangleq \Pr(\mathbf{U} + \mathbf{Z} \notin \mathcal{V})$$

$$P_{\mathcal{V}} \triangleq \Pr(\mathbf{U} + \mathbf{Z} \in \mathcal{V}),$$

where $P_{\overline{\mathcal{V}}} = 1 - P_{\mathcal{V}}$. In [20], it was shown that for lattices which are good for AWGN channel coding (27), we have that $P_{\overline{\mathcal{V}}} < \epsilon'$ for any $\epsilon' > 0$. We can write that

$$[\mathbf{U} + \mathbf{Z}] \bmod \Lambda_n = \begin{cases} \mathbf{U} + \mathbf{Z}, & P = P_{\mathcal{V}} \\ \mathbf{U} + \mathbf{Z} - Q_{\Lambda_n}(\mathbf{U} + \mathbf{Z}), & P = P_{\overline{\mathcal{V}}} \end{cases}.$$

Therefore, for sequence of lattices which are good for AWGN channel coding, we have that

$$\lim_{P_{\overline{\mathcal{V}}} \to 0} \frac{1}{n} h\left([\mathbf{U} + \mathbf{Z}] \bmod \Lambda_n\right) = \frac{1}{n} h(\mathbf{U} + \mathbf{Z}).$$

On the other hand, from the power inequality [18], we have that

$$\frac{1}{n}h(\mathbf{U}+\mathbf{Z}) \geq \frac{1}{2}\log_2\left(2^{\frac{2}{n}h(\mathbf{U})} + 2^{\frac{2}{n}h(\mathbf{Z})}\right)$$

$$= \frac{1}{2}\log_2\left(2^{\log_2\left(\frac{\kappa^2 P}{G(\Lambda_n)}\right)} + 2^{\log_2(2\pi eN)}\right)$$

$$= \frac{1}{2}\log_2\left(\frac{\kappa^2 P}{G(\Lambda_n)} + 2\pi eN\right).$$

For sequence of lattices which are good for quantization (26), we have that $G(\Lambda_n) = \frac{1}{2\pi e(1-\epsilon')}$ where $\epsilon' \to 0$, thus

$$\frac{1}{n}h(\mathbf{U}+\mathbf{Z}) \geq \frac{1}{2}\log_2\left(2\pi e\kappa^2 P(1-\epsilon') + 2\pi eN\right)$$

$$\geq \frac{1}{2}\log_2\left(2\pi eP(1-\epsilon')\right)$$

$$= \frac{1}{2}\log_2\left(2\pi eP\right) - \frac{1}{2}\log_2\left(\frac{1}{1-\epsilon'}\right)$$

The lemma follows from the definition $\epsilon \triangleq \frac{1}{2}\log_2\left(\frac{1}{1-\epsilon'}\right)$, hence $\epsilon \to 0$ as $\epsilon' \to 0$ for good lattice for quantization.

## Appendix V

## Proof of Theorem 5

In the following transmission scheme we use the lattices $\Lambda_1$ and $\Lambda_2$ with fundamental Voronoi region $\mathcal{V}_1$ and $\mathcal{V}_2$ with second moment $P_1$ and $P_2$, respectively. We further assume that the lattices are both good for quantization (26) and good for AWGN channel coding (27).

We consider the case that $P_2 = P_1 + N$. Using the canonical transmission scheme of Section IV-B with $\mathbf{V}_1 = \mathbf{0}$, $\mathbf{D}_2 = \mathbf{0}$, $\Lambda_r = \Lambda_1$, $\gamma = 1$ and $\alpha_r = \alpha_1$, the encoders send

$$\mathbf{X}_1 = [-\alpha_1\mathbf{S}_1 + \mathbf{D}_1] \bmod \Lambda_1$$

$$\mathbf{X}_2 = \mathbf{V}_2, \tag{140}$$

where $\mathbf{V}_2 \sim \mathrm{Unif}(\mathcal{V}_2)$ carries the information of user 2 and $\mathbf{D}_1 \sim \mathrm{Unif}(\mathcal{V}_1)$. Due to the dither property the transmitted signal $\mathbf{X}_1 \sim \mathrm{Unif}(\mathcal{V}_1)$. The receiver calculates $\mathbf{Y}' = [\alpha_1\mathbf{Y} - \mathbf{D}_1] \bmod \Lambda_1$. The equivalent channel is given by

$$\mathbf{Y}' = [\alpha_1(\mathbf{X}_1 + \mathbf{X}_2 + \mathbf{S}_1 + \mathbf{Z}) - \mathbf{D}_1] \bmod \Lambda_1$$

$$= [\alpha_1\mathbf{V}_2 - (1-\alpha_1)[-\alpha_1\mathbf{S}_1 + \mathbf{D}_1 - Q_{\Lambda_1}(-\alpha_1\mathbf{S}_1 + \mathbf{D}_1)] + \alpha_1\mathbf{Z} - Q_{\Lambda_1}(-\alpha_1\mathbf{S}_1 + \mathbf{D}_1)] \bmod \Lambda_1$$

$$= [\alpha_1\mathbf{V}_2 - (1-\alpha_1)\mathbf{X}_1 + \alpha_1\mathbf{Z}] \bmod \Lambda_1, \tag{141}$$

where $\mathbf{X}_1$ and $\mathbf{V}_2$ are independent. The scalar $\alpha_1$ is determined to be the optimal MMSE factor, i.e., $\alpha_1 = \frac{P_1}{P_1+N}$, furthermore we select the lattices such that $\Lambda_1 = \alpha_1\Lambda_2$, hence $\alpha_1\mathbf{V}_2 \sim \mathrm{Unif}(\mathcal{V}_1)$. Since $\mathbf{V}_2$ and $\mathbf{X}_1$ are independent, the rate achieved by user 2 is given by

$$
\begin{aligned}
R_2 &= \frac{1}{n} I(\mathbf{V}_2; \mathbf{Y}') = \frac{1}{n} \left\{ h(\mathbf{Y}') - h(\mathbf{Y}'|\mathbf{V}_2) \right\} \\
&= \frac{1}{n} \left\{ h(\mathbf{Y}') - h([(1-\alpha_1)\mathbf{X}_1 + \alpha_1 \mathbf{Z}] \bmod \Lambda_1) \right\} \\
&\geq \frac{1}{2} \log_2 \left( \frac{P_1}{G(\Lambda_1)} \right) - \frac{1}{2} \log_2 \left( 2\pi e \left( (1-\alpha_1)^2 P_1 + \alpha_1{}^2 N \right) \right) \quad (142) \\
&= \frac{1}{2} \log_2 \left( 1 + \frac{P_1}{N} \right) - \frac{1}{2} \log_2 \left( 2\pi e G(\Lambda_1) \right). \quad (143)
\end{aligned}
$$

Using lattices which are good for quantization, i.e., $G(\Lambda_1), G(\Lambda_2) \to 1/2\pi e$ as $n \to \infty$, we get that any rate

$$
R_2 \leq \frac{1}{2} \log_2 \left( 1 + \frac{P_1}{N} \right) \quad (144)
$$

is achievable. Therefore, for $P_2 = P_1 + N$ the inner bound meets the outer bound (17). For $P_2 \geq P_1 + N$, the outer bound (17) remains $\frac{1}{2} \log_2(1 + P_1/N)$, which is clearly achievable.

Now, we assume that $P_1 = P_2 + N$. We use the same transmission scheme as in (140), but with $\alpha_1 = 1$. From (141), the equivalent channel is given by

$$
\mathbf{Y}' = [\mathbf{V}_2 + \mathbf{Z}] \bmod \Lambda_1, \quad (145)
$$

We select the lattices such that $\Lambda_2 = \sqrt{\frac{P_2}{P_1}} \Lambda_1$. If we further assume that the lattices are both good for quantization (26) and for AWGN channel coding (27), from Lemma 3 the rate achieved by user 2 is given by

$$
\begin{aligned}
R_2 &= \frac{1}{n} I(\mathbf{V}_2; \mathbf{Y}') = \frac{1}{n} \left\{ h(\mathbf{Y}') - h(\mathbf{Z} \bmod \Lambda_1) \right\} \\
&\geq \frac{1}{2} \log_2 \left( 2\pi e P_1 \right) - \frac{1}{2} \log_2 \left( 2\pi e N \right) - \epsilon \\
&= \frac{1}{2} \log_2 \left( \frac{P_1}{N} \right) - \epsilon \\
&= \frac{1}{2} \log_2 \left( 1 + \frac{P_2}{N} \right) - \epsilon,
\end{aligned}
$$

where for $n \to \infty$ we have that $\epsilon \to 0$. Therefore, for $P_1 = P_2 + N$ the inner bound meets the outer bound (17). For $P_1 \geq P_2 + N$, the outer bound (17) remains $\frac{1}{2} \log_2(1 + P_2/N)$, which is also achievable.

## APPENDIX VI
## PROOF OF LEMMA 4

Clearly, it is only required to prove the achievable rate inside the upper convex envelope operation (61), since the region including the upper convex envelope may be achieved using time sharing.

We use the lattices $\Lambda_1$ and $\Lambda_2$ with fundamental Voronoi region $\mathcal{V}_1$ and $\mathcal{V}_2$ with second moment $P_1$ and $P_2$, respectively. We further assume that the lattices are both good for quantization (26) and good for AWGN channel coding (27). Using the canonical transmission scheme of Section IV-B with $\mathbf{V}_1 = \mathbf{0}$, $\mathbf{D}_2 = \mathbf{0}$, $\Lambda_r = \Lambda_1$, $\gamma = 1$ and $\alpha_r = \alpha_1$, the encoders send

$$
\begin{aligned}
\mathbf{X}_1 &= [-\alpha_1 \mathbf{S}_1 + \mathbf{D}_1] \bmod \Lambda_1 \\
\mathbf{X}_2 &= \mathbf{V}_2, \quad (146)
\end{aligned}
$$

where $\mathbf{V}_2 \sim \text{Unif}(\mathcal{V}_2)$ carries the information of user 2, and $\mathbf{D}_1 \sim \text{Unif}(\mathcal{V}_1)$, and $\Lambda_2 = \sqrt{\frac{P_2}{P_1}}\Lambda_1$. The receiver calculates $\mathbf{Y}' = [\alpha_1 \mathbf{Y} - \mathbf{D}_1] \mod \Lambda_1$. The equivalent channel is given by

$$\mathbf{Y}' = [\alpha_1(\mathbf{x}_1 + \mathbf{X}_2 + \mathbf{S}_1 + \mathbf{Z}) - \mathbf{D}_1] \mod \Lambda_1$$

$$= [\alpha_1 \mathbf{V}_2 - (1-\alpha_1)[-\alpha_1 \mathbf{S}_1 + \mathbf{D}_1 - Q_{\Lambda_1}(-\alpha_1 \mathbf{S}_1 + \mathbf{D}_1)] + \alpha_1 \mathbf{Z} - Q_{\Lambda_1}(-\alpha_1 \mathbf{S}_1 + \mathbf{D}_1)] \mod \Lambda_1$$

$$= [\alpha_1 \mathbf{V}_2 - (1-\alpha_1)\mathbf{X}_1 + \alpha_1 \mathbf{Z}] \mod \Lambda_1, \tag{147}$$

We determine the scalar $\alpha_1$ such that the second moment of $\alpha_1 \mathbf{V}_2 - (1-\alpha_1)\mathbf{X}_1 + \alpha_1 \mathbf{Z}$ will be $P_1$, i.e., $\alpha_1^2(P_2 + N) + (1-\alpha_1)^2 P1 = P1$, hence

$$\alpha_1 = \frac{2P1}{P_1 + P_2 + N} \triangleq \alpha_1^*. \tag{148}$$

Since the lattices are good for quantization (26) and good for AWGN channel coding (27), from Lemma 3 the achievable rate of user 2 is given by

$$R_2 = \frac{1}{n}I(\mathbf{V}_2; \mathbf{Y}') = \frac{1}{n}\left\{h(\mathbf{Y}') - h([(1-\alpha_1^*)\mathbf{X}_1 + \alpha_1^*\mathbf{Z}] \mod \Lambda_1)\right\}$$

$$\geq \frac{1}{n}h(\mathbf{Y}') - \frac{1}{2}\log_2\left(2\pi e((1-\alpha_1^*)^2 P_1 + \alpha_1^{*2}N)\right)$$

$$\geq \frac{1}{2}\log_2\left(2\pi e P_1\right) - \frac{1}{2}\log_2\left(2\pi e((1-\alpha_1^*)^2 P_1 + \alpha_1^{*2}N)\right) - \epsilon \tag{149}$$

$$= \frac{1}{2}\log_2\left(\frac{P_1}{\frac{P_1(P_2-P_1+N)^2 + 4P_1^2 N}{(P_1+P_2+N)^2}}\right) - \epsilon \tag{150}$$

$$= \frac{1}{2}\log_2\left(1 + \frac{4P_1 P_2}{(P_2 - P_1 + N)^2 + 4P_1 N}\right) - \epsilon. \tag{151}$$

The proof follows since for $n \to \infty$ we have that $\epsilon$ goes to zero.

# APPENDIX VII

## PROOF OF LEMMA 5

For $P_1 \leq P_2$ and fixed $P_1$, the gap $\gamma(P_1, P_2)$ is decreasing with respect to $P_2$. Therefore, $\gamma(P_1, P_2) \leq \gamma(P_1, P_1)$. In the same way, it can be shown that for $P_1 \geq P_2$ for fixed $P_2$, $\gamma(P_1, P_2) \leq \gamma(P_2, P_2)$. As a consequence, we have that

$$\gamma(P_1, P_2) \leq \gamma(P_{\min}, P_{\min}), \tag{152}$$

where $P_{\min} = \min\{P_1, P_2\}$.

Since the upper convex envelope can only decrease the gap, hence

$$\gamma(P_{\min}, P_{\min}) \leq \frac{1}{2}\log_2\left(1 + \frac{P_{\min}}{N}\right) - \frac{1}{2}\log_2\left(1 + \frac{4P_{\min}^2}{N^2 + 4P_{\min}N}\right) \tag{153}$$

$$\leq \max_{P_{\min}, N} \frac{1}{2}\log_2\left(\frac{P_{\min} + N}{N} \cdot \frac{4P_{\min}N + N^2 + 4P_{\min}^2}{N^2 + 4P_{\min}N}\right) \tag{154}$$

$$= \max_{P_{\min}, N} \frac{1}{2}\log_2\left(\frac{(P_{\min} + N)(4P_{\min} + N)}{(2P_{\min} + N)^2}\right) \tag{155}$$

$$= \max_{P_{\min}, N} \frac{1}{2}\log_2\left(\frac{(1 + P_{\min}/N)(1 + 4P_{\min}/N)}{(1 + 2P_{\min}/N)^2}\right). \tag{156}$$

The proof follows since the maximum of the function $f(x) = \frac{(1+x)(1+4x)}{(1+2x)^2}$ occurs at $x^* = 1/2$, and $f(x^*) = 9/8$.

## Appendix VIII

## Proof of Theorem 6

We first consider the case that $P_1 \leq P_2$. In this case we use the lattice $\Lambda_1$ with fundamental Voronoi region $\mathcal{V}_1$ and second moment $P_1$. We further assume that the lattice $\Lambda_1$ is good for quantization (26). Using the canonical transmission scheme of Section IV-B with $\mathbf{D}_1 = \mathbf{D}_2 = \mathbf{0}$, $\Lambda_1 = \Lambda_2 = \Lambda_r$ and $\alpha_1 = \alpha_2 = \alpha_r = 1$, the encoders send

$$\mathbf{X}_1 = [\mathbf{V}_1 - \mathbf{S}_1] \bmod \Lambda_1 \tag{157}$$

$$\mathbf{X}_2 = \mathbf{V}_2, \tag{158}$$

where $\mathbf{V}_1, \mathbf{V}_2 \sim \mathrm{Unif}(\mathcal{V}_1)$ carry the information of user 1 and user 2, respectively. The power constraints are satisfied since $\mathbf{X}_1 \sim \mathrm{Unif}(\mathcal{V}_1)$, $\mathbf{X}_2 \sim \mathrm{Unif}(\mathcal{V}_1)$ and $P_1 \leq P_2$. The receiver calculates $\mathbf{Y}' = \mathbf{Y} \bmod \Lambda_1$. The equivalent channel is given by

$$\mathbf{Y}' = [\mathbf{V}_1 + \mathbf{V}_2 + \mathbf{Z} - Q_{\Lambda_1}(\mathbf{V}_1 - \mathbf{S}_1)] \bmod \Lambda_1 \tag{159}$$

$$= [\mathbf{V}_1 + \mathbf{V}_2 + \mathbf{Z}] \bmod \Lambda_1. \tag{160}$$

The rate sum is given by

$$R_1 + R_2 = \frac{1}{n} I(\mathbf{V}_1, \mathbf{V}_2; \mathbf{Y}')$$
$$= \frac{1}{n} \left\{ h(\mathbf{Y}') - h(\mathbf{Z} \bmod \Lambda_1) \right\}$$
$$\geq \frac{1}{2} \log_2 \left( \frac{P_1}{G_n(\Lambda_1)} \right) - \frac{1}{2} \log_2 \left( 2\pi e N \right)$$
$$= \frac{1}{2} \log_2 \left( \frac{P_1}{N} \right) - \frac{1}{2} \log_2 \left( 2\pi e G_n(\Lambda_1) \right).$$

Since the lattice $\Lambda_1$ is good for quantization, i.e., $G(\Lambda_1) \to 1/2\pi e$ as $n \to \infty$, we get that any rate pair

$$R_1 + R_2 \leq \frac{1}{2} \log_2 \left( \frac{P_1}{N} \right) - o(1). \tag{161}$$

is achievable, where $o(1) \to 0$ as $P_1 \to \infty$. This achievable region coincides with the outer bound (9) for $P_1 \leq P_2$.

We now consider that case that $P_1 > P_2$. We use the lattices $\Lambda_1$ and $\Lambda_2$ with fundamental Voronoi region $\mathcal{V}_1$ and $\mathcal{V}_2$ with second moment $P_1$ and $P_2$, respectively. Using the canonical transmission scheme of Section IV-B with $\mathbf{D}_1 = \mathbf{D}_2 = \mathbf{0}$, $\Lambda_r = \Lambda_1$ and $\alpha_1 = \alpha_2 = \alpha_r = 1$, the encoders send

$$\mathbf{X}_1 = [\mathbf{V}_1 - \mathbf{S}_1] \bmod \Lambda_1 \tag{162}$$

$$\mathbf{X}_2 = \mathbf{V}_2, \tag{163}$$

where $\mathbf{V}_1 \sim \mathrm{Unif}(\mathcal{V}_1)$ and $\mathbf{V}_2 \sim \mathrm{Unif}(\mathcal{V}_2)$ carry the information of user 1 and user 2, respectively. The power constraints are satisfied since $\mathbf{X}_1 \sim \mathrm{Unif}(\mathcal{V}_1)$, $\mathbf{X}_2 \sim \mathrm{Unif}(\mathcal{V}_2)$. The receiver calculates $\mathbf{Y}' = \mathbf{Y} \bmod \Lambda_1$. Again, the equivalent channel is given by

$$\mathbf{Y}' = [\mathbf{V}_1 + \mathbf{V}_2 + \mathbf{Z} - Q_{\Lambda_1}(\mathbf{V}_1 - \mathbf{S}_1)] \bmod \Lambda_1 \tag{164}$$

$$= [\mathbf{V}_1 + \mathbf{V}_2 + \mathbf{Z}] \bmod \Lambda_1. \tag{165}$$

We use nested lattice structure [21], i.e., $\Lambda_1 \subset \Lambda_2$. We further assume that the lattices are both good for quantization (26) and good for AWGN channel coding (27). The decoder uses successive decoding to reconstruct $\mathbf{V}_1$ and $\mathbf{V}_2$ in (165). First the decoder decodes $\mathbf{V}_1$ where $\mathbf{V}_2$ acts as noise, thus user 1 can achieve the following rate

$$R_1 = \frac{1}{2}\log_2\left(\frac{P_1}{P_2 + N}\right)$$

Then, the decoder subtracts the reconstruction of $\mathbf{V}_1$ and reduces the result modulo $\Lambda_2$, in this case the equivalent channel is given by

$$\mathbf{Y}'' = [\mathbf{V}_2 + \mathbf{Z}] \bmod \Lambda_2.$$

Therefore, user 2 achieves the rate

$$R_2 = \frac{1}{2}\log_2\left(\frac{P_2}{N}\right).$$

Clearly at high SNR, i.e., for $P_1, P_2 \gg N$, this achievable rate pair coincides with the point $(R_1^c, R_2^c)$ (16). From Lemma 6, the rate pair $(0, R_2) = (0, \frac{1}{2}\log_2\left(\frac{P_2}{N}\right) - o(1))$ is also achievable at high SNR. Likewise, the point $(R_1, 0) = (0.5 \cdot \log_2(1 + P_1/N), 0)$ is achievable for any SNR. The theorem follows since the region defined by the time sharing between these three points coincides with the outer bound (9) at high SNR .

# APPENDIX IX

## PROOF OF LEMMA 7

we use the lattices $\Lambda_1$ and $\Lambda_2$ with fundamental Voronoi region $\mathcal{V}_1$ and $\mathcal{V}_2$ with second moment $P_1$ and $P_2$, respectively. We further assume that the lattices are both good for quantization (26) and good for AWGN channel coding (27), and $\Lambda_2 = \sqrt{\frac{P_2}{P_1}}\Lambda_1$. Using the canonical transmission scheme of Section IV-B with $\mathbf{D}_2 = \mathbf{0}$, $\Lambda_r = \Lambda_1$, $\gamma = 1$ and $\alpha_1 = \alpha_r$, the encoders send

$$\mathbf{X}_1 = [\mathbf{V}_1 - \alpha_1\mathbf{S}_1 + \mathbf{D}_1] \bmod \Lambda_1$$

$$\mathbf{X}_2 = \mathbf{V}_2, \tag{166}$$

where $\mathbf{V}_1 \sim \mathrm{Unif}(\mathcal{V}_1)$ and $\mathbf{V}_2 \sim \mathrm{Unif}(\mathcal{V}_2)$ are independent and carry the information of user 1 and user 2 respectively, and $\mathbf{D}_1 \sim \mathrm{Unif}(\mathcal{V}_1)$. Due to the dither quantization property, the power constraint of user 1 is satisfied, furthermore the random vectors $\mathbf{V}_1$ and $\mathbf{X}_1$ are independent. The receiver calculates $\mathbf{Y}' = [\alpha_1\mathbf{Y} - \mathbf{D}_1] \bmod \Lambda_1$.

The equivalent channel is given by

$$\mathbf{Y}' = [\alpha_1(\mathbf{X}_1 + \mathbf{X}_2 + \mathbf{S}_1 + \mathbf{Z}) - \mathbf{D}_1] \bmod \Lambda_1$$

$$= [\mathbf{V}_1 + \alpha_1\mathbf{V}_2 - (1-\alpha_1)[\mathbf{V}_1 - \alpha_1\mathbf{S}_1 + \mathbf{D}_1 - Q_{\Lambda_1}(\mathbf{V}_1 - \alpha_1\mathbf{S}_1 + \mathbf{D}_1)] + \alpha_1\mathbf{Z} - Q_{\Lambda_1}(\mathbf{V}_1 - \alpha_1\mathbf{S}_1 + \mathbf{D}_1)] \bmod \Lambda_1$$

$$= [\mathbf{V}_1 + \alpha_1\mathbf{V}_2 - (1-\alpha_1)\mathbf{X}_1 + \alpha_1\mathbf{Z}] \bmod \Lambda_1,$$

The rate achieved by user 1 is given by

$$R_1 = \frac{1}{n}I(\mathbf{V}_1; \mathbf{Y}') = \frac{1}{n}\left\{h(\mathbf{Y}') - h(\mathbf{Y}'|\mathbf{V}_1))\right\} \tag{167}$$

$$= \frac{1}{n}\left\{h(\mathbf{Y}') - h([\alpha_1\mathbf{V}_2 + (1-\alpha_1)\mathbf{X}_1 + \alpha_1\mathbf{Z}] \bmod \Lambda_1)\right\} \tag{168}$$

$$\geq \frac{1}{n}\left\{h(\mathbf{Y}') - \min\left\{\frac{1}{2}\log_2(2\pi e P_1), h(\alpha_1\mathbf{V}_2 + (1-\alpha_1)\mathbf{X}_1 + \alpha_1\mathbf{Z})\right\}\right\} \tag{169}$$

$$\geq \frac{1}{2}\log_2\left(\frac{P_1}{G(\Lambda_1)}\right) - \frac{1}{2}\log_2\left(2\pi e \cdot \min\left\{P_1, \alpha_1^2 P_2 + (1-\alpha_1)^2 P_1 + \alpha_1^2 N\right\}\right) \tag{170}$$

$$= \frac{1}{2}\log_2\left(\frac{P_1}{\min\left\{P_1, \alpha_1^2 P_2 + (1-\alpha_1)^2 P_1 + \alpha_1^2 N\right\}}\right) - \frac{1}{2}\log_2\left(2\pi e G(\Lambda_1)\right), \tag{171}$$

where (169) follows since $h(\mathbf{U} \bmod \Lambda_1) \leq \min\{\frac{n}{2}\log_2(2\pi e P_1), h(\mathbf{U})\}$ for any random vector $\mathbf{U}$; (170) follows since $\mathbf{Y}' \sim \text{Unif}(\mathcal{V}_1)$ thus $h(\mathbf{Y}') = \frac{1}{2}\log_2\left(\frac{P_1}{G_n(\Lambda_1)}\right)$, and since Gaussian distribution maximizes the entropy for fixed variance. Using lattice $\Lambda_1$ which is good for quantization, i.e., $G(\Lambda_1) \to 1/2\pi e$ as $n \to \infty$, we get that any rate

$$R_1 \leq \frac{1}{2}\log_2\left(\frac{P_1}{\min\left\{P_1, (1-\alpha_1)^2 P_1 + \alpha_1^2(N + P_2)\right\}}\right) \tag{172}$$

is achievable. Since $\mathbf{V}_1$ is reconstructed at the decoder with high probability, we can subtract $\hat{\mathbf{V}}_1$ from $\mathbf{Y}'$. i.e

$$\tilde{\mathbf{Y}} = [\mathbf{Y}' - \hat{\mathbf{V}}_1] \bmod \Lambda_1 \tag{173}$$

$$= [\alpha_1\mathbf{V}_2 - (1-\alpha_1)\mathbf{X}_1 + \alpha_1\mathbf{Z}] \bmod \Lambda_1. \tag{174}$$

In order to reconstruct $\mathbf{V}_2$, the receiver calculates $\mathbf{Y}'' = [\tilde{\mathbf{Y}}] \bmod \Lambda_r'$ where the lattice $\Lambda_r'$ satisfies $\Lambda_r' = \beta\Lambda_1$ where $\beta = \sqrt{\frac{\min\{P_1, (1-\alpha_1)^2 P_1 + \alpha_1^2(N + P_2)\}}{P_1}}$. The equivalent channel is given by

$$\mathbf{Y}'' = [\alpha_1\mathbf{V}_2 - (1-\alpha_1)\mathbf{X}_1 + \alpha_1\mathbf{Z}] \bmod \Lambda_r'. \tag{175}$$

Since the lattices $\Lambda_1$ and $\Lambda_2$ are both good for quantization (26) and good for AWGN channel coding (27), hence $\Lambda_r'$ is both good for quantization and for AWGN channel coding as well. Therefore, the rate achieved by user 2 is

given by

$$R_2 = \frac{1}{n} I(\mathbf{V}_2; \mathbf{Y}'') = \frac{1}{n} \left\{ h(\mathbf{Y}'') - h(\mathbf{Y}''|\mathbf{V}_2)) \right\} \tag{176}$$

$$= \frac{1}{n} \left\{ h(\mathbf{Y}'') - h([(1-\alpha_1)\mathbf{X}_1 + \alpha_1 \mathbf{Z}] \bmod \Lambda_r') \right\} \tag{177}$$

$$\geq \frac{1}{n} h(\mathbf{Y}'') - \frac{1}{2} \log_2 \left( 2\pi e \left( (1-\alpha_1)^2 P_1 + \alpha_1^2 N \right) \right) \tag{178}$$

$$\geq \frac{1}{2} \log_2 \left( 2\pi e \cdot \min\{P_1, (1-\alpha_1)^2 P_1 + \alpha_1^2 (P_2 + N)\} \right) - \frac{1}{2} \log_2 \left( 2\pi e \left( (1-\alpha_1)^2 P_1 + \alpha_1^2 N \right) \right) - \epsilon \tag{179}$$

$$= \frac{1}{2} \log_2 \left( \frac{\min\{P_1, (1-\alpha_1)^2 P_1 + \alpha_1^2 (P_2 + N)\}}{(1-\alpha_1)^2 P_1 + \alpha_1^2 N} \right) - \epsilon, \tag{180}$$

where (179) follows from Lemma 3 where $\epsilon \to 0$ as $n \to \infty$.

## REFERENCES

[1] T. Philosof, A. Khisti, U. Erez, and R. Zamir, "Lattice strategies for the dirty multiple access channel," in *Proceedings of IEEE International Symposium on Information Theory, Nice, France*, June 2007.

[2] M. Costa, "Writing on dirty paper," *IEEE Trans. Information Theory*, vol. IT-29, pp. 439–441, May 1983.

[3] S. Gelfand and M. S. Pinsker, "Coding for channel with random parameters," *Problemy Pered. Inform. (Problems of Inform. Trans.)*, vol. 9, No. 1, pp. 19–31, 1980.

[4] S. I. Gel'fand and M. S. Pinsker, "On Gaussian channels with random parameters," in *Abstracts of Sixth International Symposium on Information Theory, Tashkent, U.S.S.R*, Sep. 1984, pp. 247–250.

[5] U. Erez, S. Shamai, and R. Zamir, "Capacity and lattice strategies for canceling known interference," *IEEE Trans. Information Theory*, vol. IT-51, pp. 3820–3833, Nov. 2005.

[6] T. Philosof and R. Zamir, "On the rate loss of single letter characterization: The dirty multiple access channel," *arXiv:0803.1120v1, also accepted to Transaction on Information Theory*, 2008.

[7] J. Korner and K. Marton, "How to encode the modulo-two sum of binary sources," *IEEE Trans. Information Theory*, vol. IT-25, pp. 219–221, March 1979.

[8] A. Somekh-Baruch, S. S. (Shitz), and S. Verdu, "Cooperative multiple-access encoding with states available at one transmitter," *IEEE Trans. Information Theory*, vol. IT-54, pp. 4448–4469, Oct. 2008.

[9] S. Kotagiri and J. N. Laneman, "Multiple access channels with state information known at some encoders," *IEEE Trans. Information Theory*, July 2006, submitted for publication.

[10] Y. H. Kim, A. Sutivong, and S. Sigurjonsson, "Multiple user writing on dirty paper," in *Proceedings of IEEE International Symposium on Information Theory, Chicago, USA*, June 2004.

[11] T. Philosof and R. Zamir, "The rate loss of single letter characterization for the "dirty" multiple access channel," in *Proceedings of IEEE Information Theory Workshop, ITW 2008, Porto, Portugal*, May 2008.

[12] S. A. Jafar, "Capacity with causal and non-causal side information - a unified view," *IEEE Trans. Information Theory*, vol. IT-52, pp. 5468–5475, Dec. 2006.

[13] J. H. Conway and N. J. A. Sloane, *Sphere Packings, Lattices and Groups*. New York, N.Y.: Springer-Verlag, 1988.

[14] R. Zamir and M. Feder, "On lattice quantization noise," *IEEE Trans. Information Theory*, vol. IT-42, pp. 1152–1159, July 1996.

[15] U. Erez, S. Litsyn, and R. Zamir, "Lattices which are good for (almost) everything," *IEEE Trans. Information Theory*, vol. IT-51, pp. 3401–3416, Oct. 2005.

[16] C. E. Shannon, "Channels with side information at the transmitter," *IBM Journal of Research and Development*, vol. 2, pp. 289–293, Oct. 1958.

[17] S. Boyd and L. Vandenberghe, *Convex Optimization*. Cambridge: Cambridge University Press, 2004.

[18] T. M. Cover and J. A. Thomas, *Elements of Information Theory*. New York: Wiley, 1991.

[19] F. Willems, "Signalling for the Gaussian channel with side information at the transmitter," in *Proc. Int. Symp. Inform. Theory (ISIT), Sorrento, Italy*, June 2000, p. 348.

[20] U. Erez and R. Zamir, "Achieving $\frac{1}{2}\log(1+SNR)$ on the AWGN channel with lattice encoding and decoding," *IEEE Trans. Information Theory*, vol. 50, pp. 2293–2314, Oct. 2004.

[21] R. Zamir, S. Shamai, and U. Erez, "Nested linear/lattice codes for structured multiterminal binning," *IEEE Trans. Information Theory*, vol. IT-48, pp. 1250–1276, June 2002.